# Auditing bipartite motif interpretations: a worked example with conservation checks and open-path decomposition

**Tengfei Shao**

Global Education Center, Waseda University, Tokyo, Japan

ORCID: 0000-0003-3089-5696

Corresponding author: tengfei.shao@toki.waseda.jp

# Auditing bipartite motif interpretations: a worked example with conservation checks and open-path decomposition

*Tengfei Shao, Global Education Center, Waseda University, Tokyo, Japan;*
*tengfei.shao@toki.waseda.jp*

## Abstract

**Background.** Motif profiles of bipartite agent-object networks, such as tourist-site visits and customer-item transactions, are read as evidence about structural roles and about differences between networks, often without asking what the two degree sequences already fix. In a simple bipartite graph the induced k-fan count on one node type is a sum of degree combinations, so it has zero variance under a null that preserves both degree sequences.

**Methods.** We apply this known result to a reconstructed tourism rating network of 17 tourists, 80 sites and 637 edges, the sole inferential worked example, and, as a provenance-limited illustration, to published motif-instance aggregates over 36 monthly luxury customer-item networks. Six induced type-preserving motif classes are enumerated exactly. The non-fan statistics are scored against a hard bipartite configuration null sampled by proposal-counted double edge swaps, in two implementations, with synthetic positive and degree-matched negative controls.

**Results.** The four fan classes are exact functions of the degree sequences. In the tourism network the raw fan counts and the size-3 two-fan ratio, 84.7 % fan-out, restate those sequences; size-4 within-size shares do not. The luxury aggregates support no such reading: their published item-side counts require at least 89,502 customer-item edges against 26,451 reported transactions, failing the edge bound the identity implies, so their 99.8 % fan-in is reported as a published descriptive value only, with no determinacy verdict. Of the two size-4 statistics free to move, the four-cycle count is degree-consistent marginally (z about +1.0) and the open path is deficient (z about −5.8) by −2,243 instances, 2.4 % of the null mean. That deficit is extreme relative to the sampled hard-degree null for this observed graph and survives every leave-one-tourist-out re-run (z −4.5 to −8.1). An exact identity splits it at the point estimate into 64.4 % mixing and 35.6 % four-cycle, but that is not an attribution: the observed mixing term lies below all 500 samples drawn, so any conditional expectation there is extrapolation; the components are almost collinear under the null (r = 0.968) and the observed pair lies far outside the joint null cloud (Mahalanobis distance 17.3, not a test). It is also unstable: 36.6 to 116.5 % under leave-one-tourist-out deletion, 15.1 % without the two most active tourists. The deficit weakens as the rating threshold rises, from z = −6.1 to −1.4. The swap-counting correction shifts z by at most 0.27, below the 0.66 seed-to-seed gap at the same chain length.

**Discussion.** The deficit is therefore extreme relative to that sampled null, while its class-level interpretation is undetermined and unstable. We give a four-step pre-interpretation check and a reference implementation. The evidence is one worked example plus one published illustration, not a population estimate.

## Background

A summary statistic is only informative if some of its variation is free to differ from what the input already fixes. Analysis pipelines fail quietly when this condition is not checked, because the failure does not look like an error: the code runs, the number is reported, and the interpretation follows. Recent cautionary work in this journal has documented the pattern in other settings: a target encoding computed before resampling leaks label information into the out-of-bag sample, so an apparently validated performance estimate is partly a restatement of the training labels (Smith et al., 2024), and a structured audit of a hundred deep learning papers found that reported results frequently cannot be reproduced from what the papers themselves describe (Ahmed et al., 2025). Together these studies illustrate related audit patterns: each identifies a hidden dependency in an apparently routine analytical step. Bipartite agent-object interaction networks, in which agents such as tourists or customers are linked to objects such as sightseeing sites or items, are summarised in a comparable way. In the analyses re-examined here, published and unpublished, the counts of small induced subgraphs, the motif profile, were read as evidence about structural roles within a network and about differences between networks, following a reading that is standard in the wider motif literature, where a significance profile is compared across networks (Milo et al., 2002; Milo et al., 2004), and that is applied to bipartite networks in ecology, where motif positions are interpreted as species roles (Baker et al., 2015; Simmons et al., 2019a). Motif profiles are read in tourism and in recommendation settings as well, although on one-mode attraction-to-attraction mobility networks (Park and Zhong, 2022; Ding et al., 2024), on a single typed subgraph class inside a bipartite growth model (Hernández and González-Martel, 2017), or as a predictive architecture rather than an interpreted profile (Zhang et al., 2025). What such a profile adds to the two degree sequences was not asked in those analyses before the profile was interpreted.

For one family of motif classes the answer is available in closed form. In a simple bipartite graph the number of induced k-fans centred on one side equals $\Sigma_v C(d_v, k)$ summed over the nodes of that side, so under a null that fixes both degree sequences a fan count cannot vary at all. The consequence is not that fan counts are wrong but that they are not evidence of anything the degree sequences do not already carry: a z-score, an enrichment factor or a significance profile computed for a fan class against a hard degree-preserving null has a degenerate reference distribution, and a cross-network difference in a raw fan count is a difference between two degree sequences written in other units. The same holds for the size-3 fan-out to fan-in ratio, whose two components are both fans. It does not hold for within-size shares at size 4, whose denominator also contains the variable four-cycle and open-path counts. Here the fan classes also carry most of the counted instances, so this is not a corner case of the vocabulary but the part of it that carries most of the reported numbers.

The identity and its consequence for significance profiles are known. Wegner (2014), "Motif Conservation Laws for the Configuration Model", gives the star identity for orders $k = 2$ and 3, derives the conservation of the corresponding counts under the configuration model, draws the degrees-of-freedom consequence for significance profiles, and notes that the extension to different edge and node types is straightforward (Wegner, 2014). We take this as prior art and claim no theorem. The surrounding machinery is likewise mature. Exact and approximate counters for bipartite motifs and for graphlets are available as packages (Hočevar and Demšar, 2014; Simmons et al., 2019b), and nulls that hold one or both degree sequences fixed, whether by direct randomisation or by maximum-entropy construction, are established practice in bipartite analysis and in the extraction of statistically validated projections (Saracco et al., 2017; Neal, Domagalski and Sagan, 2021; Neal et al., 2024); statistical models for motif counts have also been developed (Picard et al., 2008; Ouadah, Latouche and Robin, 2022). What is available is therefore the theorem, the counters and the samplers, each in a separate literature.

The gap is operational rather than theorem-level. One published analysis of the luxury data reports per-class fan counts, and a companion paper per-class return statistics (Shao, Ieiri and Takahashi, 2025a); from the counts a size-3 orientation of 99.8 % fan-in follows, without first asking which part the two degree sequences already fix, and they report non-conserved statistics as totals rather than decomposing them (Shao, Ieiri and Takahashi, 2025b). The tourism size-3 profile of 84.7 % fan-out against 15.3 % fan-in, and the cross-dataset contrast that sets the two profiles side by side, were not reported in the 2021 tourism study (Shao, Ieiri and Hishiyama, 2021) or in the 2024 cross-dataset paper (Shao, Ieiri and Hishiyama, 2024); both were computed in the author's withdrawn manuscript from the 2021 network and the published luxury counts. None of that work, published or withdrawn, asked which part of the profile the degree sequences fix. The present note recomputes the profile from the same matrix and withdraws the interpretation the withdrawn manuscript placed on it. The same omission has a second half. For the statistics that are not degree-determined, a total departure from a null is reported as a single quantity, although at size 4 the open path (P4) and the four-cycle (C4) are linked to the degree–degree mixing term (the mixing term below) by the exact relation $P4 = \Sigma_E (d_a - 1)(d_o - 1) - 4 \cdot C4$, so a departure can be decomposed into its parts before interpretation, and the decomposition itself checked for identifiability and stability.

We therefore draw on one reconstructed interaction network re-analysed under a hard null, plus one published aggregate profile used as an illustration, and two synthetic controls under a corrected proposal-counted null, and we report what survives. The tourism network is reconstructed from the source study as 17 tourists, 80 sites and 637 edges, and it is the only inferential worked example in this note; the luxury side enters only as published per-class aggregates over 36 monthly customer-item networks covering 26,451 transactions, because the underlying records are proprietary and were not accessed, and those aggregates are internally inconsistent with the transaction total reported for the same networks in a way this note cannot resolve from published figures (Results), so the luxury side illustrates the identity and carries no determinacy verdict of its own. For the tourism network the non-fan statistics are scored against a hard bipartite degree-preserving null sampled by proposal-counted double edge swaps, with synthetic positive and degree-matched negative controls, and the null analysis is replicated in a second implementation, attrimotif 1.1.1 (Shao, 2026). Two self-corrections are disclosed rather than absorbed. An earlier manuscript built on that interpretation was rejected after review and was never made public; the present paper reuses its code and its data facts, not its headline. Separately, the sampler in our public code deposit counts accepted swaps instead of proposals, which biases the sampled ensemble; the corrected rule leaves the conclusions unchanged (Results, Limitation 6), but the deposit is affected and we say so.

This study asks a single measurement question, stated as RQ0 in Methods: when bipartite motif analyses report dominant, enriched, or cross-network-different fan motifs, what is fixed by the degree sequences? The primary contribution is not a new conservation theorem and not a new motif framework. It is a demonstrated checklist that prevents degree-determined counts from being interpreted as residual local structure, together with a bounded account of the one statistic that survives that diagnostic. That account is itself part of the result: the surviving deficit is robust to which tourist is dropped, but the class-level attribution of the deficit is not identifiable from the point split and does not hold up under leave-one-tourist-out deletion or under a change in the rating threshold that defines an edge. The audit's value is that it shows this before an interpretation is published.

The boundary of that contribution is stated here in five layers, so that no part of it is read as more than it is. (i) Known mathematics. The star conservation identity under the configuration model, and its consequence for the degrees of freedom available to a significance profile, are Wegner's (Wegner, 2014); the size-4 relation $P4 = M - 4 \cdot C4$ is elementary counting on a simple bipartite graph. Neither is claimed as a result of this note. (ii) Functionality contributed by attrimotif (Shao, 2026). The *is_degree_determined* guard and the proposal-counted double-edge-swap sampler are described in a separate software manuscript; they are used here rather than introduced here. (iii) Results reused from prior papers. The tourism rating matrix and its node counts come from the data-collection study and the 2021 analysis that reused it (Shao, Ieiri and Hishiyama, 2021), and the per-class luxury counts are taken from the published tables of the 2025 papers (Shao, Ieiri and Takahashi, 2025a; Shao, Ieiri and Takahashi, 2025b). (iv) Analyses performed only here. The reconstruction of the tourism edge list and its per-class induced census; the hard degree-preserving null tests of the non-fan statistics; the exact term-by-term decomposition of the open-path deficit; the joint-null, leave-one-tourist-out and rating-threshold instability analyses; and the sampler counting-rule correction with its sensitivity series. (v) The positive lesson. On this network the one statistic that survives the determinacy screen does depart from the null, and that departure cannot be assigned to either component of its own exact decomposition. The screen produces both findings, and an interpretation placed before the screen would have reported neither.

## Related work and theoretical background

### *Motif counts and what a null model fixes*

Motif analysis proceeds in two steps: enumerate the induced subgraphs of a given size and type signature, then compare that census against an ensemble in which some features of the data are held fixed. Enumeration is a solved problem, in general graphs through combinatorial graphlet counting (Hočevar and Demšar, 2014) and in bipartite ecological networks through dedicated counting software (Simmons et al., 2019b). The interpretive weight therefore falls on the second step, on which features the ensemble fixes.

For bipartite graphs, part of that answer is a counting identity rather than an empirical matter: the number of induced k-fans centred on one node type is the sum of degree combinations over that type, $\Sigma_v C(d_v, k)$. Wegner (2014) states this conservation constraint for the configuration model at $k = 2$ and $k = 3$, draws its consequence for the degrees of freedom available to a significance profile, and observes that the extension to higher-order stars and to different edge and node types is straightforward (Wegner, 2014). The size-3 and size-4 fan classes used in applied bipartite work lie inside that statement, which we cite as established theory: the present paper contributes no theorem and treats the fan classes as a diagnostic, not as a hypothesis to be tested.

What follows from the identity depends on the null. Under a degree-preserving null in the hard sense, where both degree sequences are held fixed edge by edge, every fan count is constant across the ensemble and its null standard deviation is exactly zero; a z-score for such a class is undefined rather than merely small. Under a soft null, which fixes the degrees only in expectation, the same counts retain variance and admit a test. The constraint is one-sided per class: a fan-out count is a function of the agent degree sequence alone and a fan-in count of the object degree sequence alone, so even a null that fixes one margin and leaves the other free holds the fan count on the fixed side exactly, and no test is restored for it. Ouadah, Latouche and Robin (2022) organise bipartite motif testing around exactly those star classes, whose counts carry the degree information, under a soft model (Ouadah, Latouche and Robin, 2022), and the bipartite null-model literature separates model families by whether the margins are fixed exactly or in expectation (Neal, Domagalski and Sagan, 2021; Neal et al., 2024). One motif class is thus testable under one null and vacuous under another, which makes naming the null load-bearing rather than a methodological footnote.

### *Bipartite motifs in application domains*

The applied inferential tradition in bipartite settings validates a different quantity from the one on which profile interpretations rest. Saracco et al. (2017) use the bipartite configuration model, a soft maximum-entropy null, to decide which pairwise co-occurrences, that is, V-motifs joining two nodes of the same type through a shared neighbour, survive in a monopartite projection (Saracco et al., 2017): a pairwise co-occurrence under a soft null, not the per-class census of induced three- and four-node motifs under a hard one. Applied work reporting fan-dominated profiles therefore has an inferential tradition available to it, but not one aimed at the statistic being interpreted, and in the work re-examined here, published and withdrawn, it is that census, not a projection, that carries the substantive readings. Outside our own work the nearest precedents for reading a per-class bipartite census as structural evidence are in ecology (Baker et al., 2015; Simmons et al., 2019a), most recently across sixty plant-pollinator networks in which subgraph positions are read as ecological roles against a benchmark (Lanuza, Allen-Perkins and Bartomeus, 2023), while the tourism and recommendation uses of motif counts are on one-mode mobility networks (Park and Zhong, 2022; Ding et al., 2024), on a single typed class inside a growth model (Hernández and González-Martel, 2017), or predictive rather than interpretive (Zhang et al., 2025).

At the target venue the nearest precedents are generic rather than topical: cautionary results that pair a controlled demonstration with a practitioner remedy (Smith et al., 2024; Ahmed et al., 2025), the statistical bar such a claim must meet in metamorphic testing of stochastic convolutional networks (Rehman and Izurieta, 2025), attributed network generation with an explicit degree-distribution component (Uludağlı and Oğuz, 2024), and, as the only bipartite paper in the recent window, a visualisation tool (Garcia-Algarra, 2026).

### *The analyses being re-examined, including our own*

The analyses re-read here are the author's own, and what each of them actually reports differs from what the re-analysis takes from it. The 2021 tourism study is the source of the network: a rating instrument covering 17 tourists and 80 sightseeing spots, analysed by enumerating induced subgraphs with ESU and reading the highest-frequency motif templates as multiple clusters of spots (Shao, Ieiri and Hishiyama, 2021); binarising its matrix at a rating above zero gives the 637 edges analysed here, and the study reports no per-class size-3 census. Of the two luxury studies, one reports per-class motif-instance counts (Shao, Ieiri and Takahashi, 2025b) and the other per-class return statistics (Shao, Ieiri and Takahashi, 2025a), both pooled over 36 monthly customer-item networks; from the counts, the opposite orientation, 99.8 % fan-in at size 3, follows (Shao, Ieiri and Takahashi, 2025b). The 2024 paper proposes a clustering model that combines network motifs with locally determined keywords and demonstrates it on the tourism and the luxury case; it reports neither class proportions nor a normalised entropy (Shao, Ieiri and Hishiyama, 2024). The tourism size-3 profile of 84.7 % fan-out against 15.3 % fan-in, and the entropy-based contrast between the two profiles, were computed in the author's withdrawn manuscript from the 2021 network and the published luxury counts, and were never published. None of this work scored any class against a degree-preserving null; on the interpretation the withdrawn manuscript placed on the contrast, see the Background and the Discussion.

The boundary against each of them can be stated in one line, with the detail in the corresponding rows of Tables 1 to 3. The 2021 tourism study enumerated motifs to discover clusters of spots and used no null model; this note adds the per-class size-3 and size-4 census on the same network, the degree-preserving null and the determinacy verdict, and changes none of its published results. The two luxury studies reported per-class instance counts and per-class return statistics pooled over 36 monthly networks, screened not against a degree-preserving null but against an edge-count-preserving random baseline that does not preserve degrees, that is, a degree-destroying null, with a significance filter applied to the resulting frequencies; this note adds the counting identity that the fan counts restate on each side, and a comparability caveat, because no edge list was available to it and the published item-side counts cannot be reconciled with the transaction total reported for the same networks (Results), so the luxury side enters as an illustration of the identity rather than as a second worked dataset. The 2024 cross-dataset paper demonstrated one clustering model on both cases and drew no profile-level contrast; the reading of the two profiles as a contrast between interaction regimes was made in the author's withdrawn manuscript, and this note withdraws that reading of the fan component and replaces it with the degree-sequence account. The software paper describes the counters, the sampler and the *is_degree_determined* guard; this note is the worked example behind that guard and supplies the open-path counter and the decomposition the package omits. Relative to all five, what is added here is the hard-null scoring of the non-fan statistics on a real network, and the finding that the decomposition of the one surviving departure does not identify an attribution.

The same code and data supported an earlier manuscript by the author, submitted to and rejected by Scientific Reports in 2026, whose title and cross-dataset framing used the fan-out versus fan-in contrast as the headline; that manuscript, and not any of the published papers, is where the tourism size-3 profile and the entropy comparison between the two profiles were computed, and its final version already stated in Methods that the fan counts are exact functions of the degree sequences. The present note withdraws the cross-dataset interpretation and adds the audit. That manuscript was never public and no preprint exists; it is disclosed here, and to the editor, as a prior attempt on overlapping data. Only its verified data facts and its code are re-used.

### *Software and reproducibility precedent*

The boundary with our own software paper can be stated in function names. attrimotif v1.1.1 (Shao, 2026) provides exact size-3 and size-4 fan counts, a four-cycle count, a proposal-counted sampler for double edge swaps, a null test accepting a custom statistic, the attribute re-attachment routines, and a guard, *is_degree_determined*, whose registry is exactly the four fan classes. It implements no open path (P4) counter and no mixing term, and ships only synthetic generators. The nulls of record use attrimotif 1.1.1, so the sampler rule of record is the proposal-counted one rather than the accepted-count rule shipped in our own earlier public deposit, a bias we found and quantify here; the sensitivity rows use the named local, deposit and legacy implementations instead.

*Nearest-neighbour comparison*

Tables 1 to 3 place the paper against its nearest neighbours in theory, in application, in our own record, and in venue genre.

**Table 1.** Nearest neighbours in theory and application: what each asked, on what data, with what method and validation, and what this note adds.

| Study | Research question | Data | Method | Outcome | Validation | Contribution | What this paper adds |
|---|---|---|---|---|---|---|---|
| Wegner (2014) | Are motif counts conserved under the configuration model? | Theory | Analytic derivation | Star counts are functions of the degree sequence; significance profiles lose degrees of freedom | Proof | Theorem | Nothing. We cite it and do not extend it |
| Ouadah, Latouche and Robin (2022) | How can bipartite motif counts be tested? | Two ecological bipartite networks: plant-pollinator (546 plants, 1,044 insects) and seed dispersal (207 plants, 110 dispersers), plus simulations | Tests for bipartite motif counts under a soft model, in which the star frequencies enter as degree-heterogeneity information | Star frequencies carry the degree information and are used to condition the tests, which target the non-star motif classes | Asymptotic theory | Statistical method | Under a hard null the star classes themselves have zero variance and are not testable, so the information they carry cannot be recovered as a test of them |
| Saracco et al. (2017) | Which pairwise co-occurrences are statistically validated? | World Trade Web (146 countries by 1,131 products, 1995 to 2010) and MovieLens 100K (943 users by 1,559 movies) | Bipartite configuration model, soft null, V-motif tests | Validated monopartite projection | Maximum-entropy ensemble | Applied inferential tool | The applied line tests a projection statistic, not the per-class census |
| Neal et al. (2021); Neal et al. (2024) (Neal, Domagalski and Sagan, 2021; Neal et al., 2024) | Which projected co-occurrences survive a null that fixes both degree sequences exactly, and how do hard and soft bipartite null families differ? | 2021: the GaWC 2000 city by firm network (196 by 100) plus four simulation studies. 2024: a review, with illustrative rather than analysed matrices | Fixed degree sequence model, a hard both-side null, plus a review of hard versus soft bipartite null families | A validated projection backbone; families separated by exactly fixed versus in-expectation margins | Simulation and family comparison | Null-model family and backbone extraction | No per-class motif audit is performed there; we add the per-class induced census under the same hard family, in which four of the six classes are degenerate |

**Table 2.** Nearest neighbours in venue: recent PeerJ Computer Science articles whose audit pattern this note follows, in the same columns.

| Study | Research question | Data | Method | Outcome | Validation | Contribution | What this paper adds |
|---|---|---|---|---|---|---|---|
| *peerj-cs.2445* (Smith et al., 2024) | Does target encoding leak into out-of-bag samples? | Simulation only | Controlled demonstration | Negative result with a remedy | Simulated ground truth | Cautionary methodology | Same genre, moved to real bipartite networks and a hard null |
| *peerj-cs.2618* (Ahmed et al., 2025) | Are published deep-learning methods reproducible? | One hundred publications | Structured audit | Reported deficiencies | Coding protocol | Negative meta-research | We audit the identifiability of a statistic, not reporting quality |
| *peerj-cs.2658* (Rehman and Izurieta, 2025) | Can bugs be detected in stochastic convolutional neural networks? | Two CNNs and their mutants | Metamorphic relations with several statistical tests | A validation procedure | Mutation ground truth | Method and validation | Rigour here comes from an exact identity, a corrected sampler, a cross-implementation check |
| *peerj-cs.2483* (Uludağlı and Oğuz, 2024) | How to generate attributed networks with communities? | Synthetic plus one real network | Property conformance against competing generators | A generator | Comparison to baselines | Method | We ask how degree constraints limit interpretation; we generate nothing |
| *peerj-cs.3526* (Garcia-Algarra, 2026) | How to make bipartite plots readable? | Ecological networks | k-core reordering, tool demonstration | A visualisation tool | Demonstration | Tool | The only direct bipartite neighbour at this venue, and not inferential |

**Table 3.** The author's own analyses and software re-examined in this note, in the same columns.

| Study | Research question | Data | Method | Outcome | Validation | Contribution | What this paper adds |
|---|---|---|---|---|---|---|---|
| Shao et al. (2021), tourism (Shao, Ieiri and Hishiyama, 2021) | Which clusters of sightseeing spots can motif enumeration discover? | Tourism rating instrument, 17 × 80; its matrix yields the 637 edges analysed here | ESU enumeration, highest-frequency motif templates read as clusters; no per-class census, no null model | Multiple clusters of spots | Tourist-satisfaction experiment | Applied clustering finding | The network itself, plus the per-class census, the degree-preserving null and the determinacy verdict it did not report |
| Shao et al. (2025), luxury (Shao, Ieiri and Takahashi, 2025a; Shao, Ieiri and Takahashi, 2025b) | Do motif classes differ in transaction outcomes? | 36 monthly networks, 26,451 transactions | Per-class counts (JIP 33:9) and return statistics (JIP 33:219); the counts are screened against an edge-count-preserving random baseline that does not preserve degrees, a degree-destroying null, with a significance filter, not against a degree-preserving null | Fan-in 99.8 % | Significance filter against a degree-destroying random baseline | Applied finding | The counting identity behind the fan counts, plus a comparability caveat on pooled monthly networks and on published aggregates that do not satisfy the edge bound the identity implies; an illustration, not a second worked dataset |
| Shao et al. (2024), cross-dataset (Shao, Ieiri and Hishiyama, 2024) | Can one clustering model serve a tourism and a transaction network? | The tourism and luxury case studies | Clustering model combining motifs with locally determined keywords; no per-class profile comparison, no motif significance test reported (a clustering model) | One model demonstrated on both cases | Cluster counts and running time against the earlier method | Cross-dataset method demonstration | The comparison design, and the degree-sequence account of the fan contrast the withdrawn manuscript drew from these sources; two datasets cannot separate domain from dataset |
| attrimotif v1.1.1 (Shao, 2026) | Software for typed motif counting and null tests | Synthetic generators | Package with unit tests | *is_degree_determined*, proposal-counted sampler, custom-statistic null test | Software testing | Reference implementation | The worked example behind that guard, plus the open-path decomposition it omits |

# Materials & Methods

## Study design and inferential target

The study is a retrospective measurement audit of two existing data sources and two synthetic controls. It is not a new data collection, not a causal study, and not an algorithm comparison; its object is the interpretation attached to published motif profiles. Causal effects, industry comparison, behavioural prediction and any new theorem are explicitly not targets.

The estimand is stated as follows. For each motif statistic, the part of its observed value that is fixed by the two degree sequences, and the distribution of the residual under a null that holds both degree sequences exactly. There is no superpopulation: the two data sources are the objects of exhaustive enumeration, so every result below is a worked example and not a population estimate.

## Research questions

The frozen scope admits one core question, split into five named diagnostic sub-questions. No hypotheses are set: the fan identity is a theorem rather than a hypothesis under test, the two datasets share no common sampling unit, and the open-path analysis is a retrospective reanalysis. Table 4 sets out the five sub-questions with their sources, units, estimands, evidentiary directions and status.

> **RQ0. To what extent are apparent fan-motif dominance, hard-null significance, and cross-dataset contrasts in two real bipartite interaction datasets determined by their degree sequences, and what remains interpretable after the degree-determined components are removed?**

**Table 4.** Research-question matrix: the five diagnostic sub-questions (A to E) under RQ0, with source, unit, estimand and comparison, the evidentiary direction that supports or weakens each reading, method and output, and status.

| Sub-question | Source | Unit | Estimand and comparison | Supports it / weakens or refutes it | Method and output | Status |
|---|---|---|---|---|---|---|
| **A. Degree determinacy.** Are the fan classes exact functions of the degree sequences, and does the hard null leave them any variance? | Wegner 2014 | Simple bipartite graph; four fan classes | Σ C(d,2) and Σ C(d,3); the null variance of each fan count | Supports: enumeration equals the identity, null sd = 0. Weakens: non-zero null variance, which would indicate an implementation fault and could not refute a theorem | Enumeration, analytic cross-check, null model; Fig. 1, Table 6 | Theorem-driven diagnostic, not a test |
| **B. Cross-dataset contrast.** Does the fan-out versus fan-in contrast survive once both degree sequences are held fixed? | Follows from A | Tourism snapshot; luxury aggregates over 36 monthly networks | Raw fan counts, the size-3 fan-out to fan-in ratio, within-size class shares, normalised entropy, across-class association at size 4 | Supports: the raw fan counts on each side, and the size-3 fan-out to fan-in ratio whose two components are both fans, are restated by the degree sequences. Weakens: a fan count not so restated. Not claimed: within-size shares at size 4, whose denominator also contains the variable four-cycle and open-path counts | Descriptive statistics with identity attribution; association as effect size only, four classes, permutation floor 0.0417; Fig. 2, Table 5 | Descriptive, explicitly not a test |
| **C. Residual and decomposition.** For the size-4 statistics that are not degree-determined, what departs from the hard null and how does it decompose? | Exact identity | Network-level statistics of the tourism network | Overlap (C4), open path (P4) and the mixing term against the proposal-counted null ensemble, with edge degree assortativity reported as a descriptive re-expression of the mixing term | Supports: four-cycle consistent with the null marginally, open path deficient under every leave-one-tourist-out re-run (z −4.45 to −8.11), and a point split of 64.4 % mixing against 35.6 % four-cycle. Does not support: that split as an attribution, because the two components are almost collinear under the null (r = 0.968), the observed mixing term lies below all 500 | Degree-preserving null plus term-by-term decomposition, the joint null geometry of the two components, leave-one-tourist-out and threshold re-runs; Fig. 3, Fig. 4, Fig. 5, Table 7, Table 8, Table 9 | Retrospective diagnostic, not pre-registered; the joint-null, deletion and threshold analyses are checks on the decomposition rather than part of the analysis plan |

| Sub-question | Source | Unit | Estimand and comparison | Supports it / weakens or refutes it | Method and output | Status |
|---|---|---|---|---|---|---|
| | | | | samples drawn so the ensemble supplies no conditional null distribution at the observed value and the joint position is reported as a Mahalanobis distance of 17.3 rather than as a test, the share runs from 36.6 to 116.5 % under leave-one-tourist-out deletion and reaches 15.1 % without the two most active tourists, and the deficit itself falls to $z = -1.43$ at the highest rating threshold | | |
| **D. Sampler robustness.** Does the swap-counting rule change any of these conclusions? | Sampler correction | As in C | Counting rule (accepted-count versus proposal-counted) × chain length (12, 60, 200 and 500 × $\lvert E \rvert$) × implementation (local versus attrimotif 1.1.1) | Supports: z shifts of at most 0.27 in z, whether measured against the record run or across the four chain lengths, and smaller than the 0.66 seed-to-seed difference at the same chain length, computed from the unrounded z values (rounded values give 0.67); mixing share moving from 66.2 % to 64.4 %. Weakens: any z crossing an interpretive boundary | Sensitivity re-runs, cross-implementation recomputation; Fig. 6, Table 10 | Robustness and disclosure |
| **E. Attribute re-attachment.** Does attaching per-instance attribute distributions add information beyond typed enumeration on these data? | Attribute workflow, reported as a negative result | Motif instances on the synthetic panel | Full per-class distribution D(c) against count, mean and 99th percentile summaries | Supports: on the synthetic panel the distribution reveals a planted maximum of 6.48 against 0.38; on the real side only published per-class aggregates exist, so a comparison of that kind is not defined there. | Synthetic unit test plus the published real-data aggregates; Fig. 7 | Exploratory, negative conclusion |

| Sub-question | Source | Unit | Estimand and comparison | Supports it / weakens or refutes it | Method and output | Status |
|---|---|---|---|---|---|---|
| | | | | Weakens: real-data structure that only Φ reveals, which these data cannot show | | |

The tested family is declared as {C4, P4, mixing term}, three statistics carrying two degrees of freedom. The identity P4 = M − 4·C4 holds exactly on the observed graph and on every graph in the ensemble, so the third statistic is a fixed linear function of the other two and the family has rank two rather than three; the three z-scores are not three independent readings of the network. The edge degree assortativity coefficient is an affine function of the mixing term on a graph with fixed degree sequences, so it is reported as a descriptive re-expression of that term and is not tested as a separate statistic. The four fan classes have zero null variance by construction and are excluded, since no test is defined for a zero-variance statistic and there is nothing for a correction to act on. The two primary statistics of sub-question C were fixed before the reanalysis began. Uncorrected z-scores and null quantiles are reported; at a family size of three no standard correction changes the reading, and a correction applied at family size three, Holm included, is conservative here for the same reason, because the effective family size is two. There is no model selection.

## Data sources, network construction, and provenance

**Tourism network.** The rating matrix was collected in a Kyoto experiment with 17 tourists and 80 sightseeing spots using a walk-rally application, and is reported in Ieiri, Nakajima and Hishiyama (2018); the 2021 study cites that paper as its reference [23] for the origin of the data and is, like this note, a secondary user of it (Shao, Ieiri and Hishiyama, 2021). Evaluations were collected for the spots a tourist actually visited, so a rating of 0 in the matrix means that the spot was not visited and no evaluation was recorded, and an edge, that is, a rating greater than 0, is a visited-and-rated spot. Agents are the 17 tourists and objects are the 80 sightseeing sites; an edge is placed where the rating is greater than zero, giving 637 edges. Rating intensity enters only through this binarisation and never enters a structural null, so the graph is simple, unweighted and bipartite. The node boundary is the fixed roster of the source instrument (Shao, Ieiri and Hishiyama, 2021); a site with no positive rating stays in the node set with degree zero and contributes zero to every $\Sigma$ C(d,k) term, so no node is dropped and no value is imputed. No one-mode projection is taken at any point: every count below is an induced, type-preserving subgraph count on the bipartite graph itself. The edge list is anonymised (agents labelled T1 to T17, no personal identifiers).

**Luxury aggregates.** The luxury side contributes published per-class motif-instance counts summed over 36 monthly customer × item networks covering 26,451 transactions after outlier removal (Shao, Ieiri and Takahashi, 2025b). Per-month node and edge cardinalities are not reported in the source and are not reconstructed here; the outlier rule and the missing-data handling belong to the source study and are not available in the present reanalysis; the underlying transaction records are proprietary and were not accessed. Motif counts do not aggregate linearly across 36 independently degree-sequenced monthly networks, so the luxury profile is not an observation symmetric to the tourism snapshot; the two are placed side by side and never tested against each other. The published item-side counts also imply a minimum customer-item edge count that the reported transaction total does not reach (Results), which cannot be resolved from published figures, so the luxury side is used as an illustration of the counting identity and no determinacy verdict is drawn from it. Provenance is summarised in Table 5.

### Motif classes and the degree-determinacy criterion

Six induced, type-preserving classes are enumerated: at size 3, fan-out (3-A, one agent and two objects) and fan-in (3-B, two agents and one object); at size 4, fan-out (4-A), overlap (4-B, the 2×2 biclique, that is, the four-cycle count, written C4; called four-cycle below), open path (4-C, written P4) and fan-in (4-D). A fan motif is a star centred on one side.

The criterion used throughout is stated as follows. A motif class is degree-determined if its induced count can be written as a function of the two degree sequences alone. For the fans this function is $\Sigma_v C(d_v, k)$.

Wegner states the corresponding conservation law for $k = 2$ and 3 (Wegner, 2014), for non-induced subgraph counts in the unipartite configuration model; in a bipartite graph with type-preserving fans the induced and non-induced counts coincide, because no object-object edge can exist, so the identity applies to the induced census used here. The present study cites that result and claims no theorem of its own.

Enumeration follows the definitions directly: fan counts as combinations over each node's neighbour set, equal by construction to $\Sigma C(d,2)$ and $\Sigma C(d,3)$; the four-cycle count as the number of 2×2 bicliques, summing $C(c,2)$ over the common-neighbourhood size c of every agent pair with $c \geq 2$; and the open-path count as the sum over agent pairs of $c \cdot ((d_i - c) + (d_j - c))$, which counts the induced configurations with two agents, two objects and exactly three edges. The degree-determinacy guard is attrimotif's *is_degree_determined* routine (Shao, 2026), used here and not a contribution of this study. The agent and object roles are defined by the researcher rather than given by the data, and only four classes exist at size 4; these are the principal sources of uncertainty in the descriptive comparison, and no interval is reported for an exhaustive census.

### Null model, degree constraints, and the corrected sampler

The null family is the hard bipartite configuration null, in which both degree sequences are preserved pointwise. It is realised by double edge swaps mapping (a1,o1) and (a2,o2) to (a1,o2) and (a2,o1), applied when the two agents differ, the two objects differ, and neither target edge is already present. Every conclusion here is limited to this family; soft and maximum-entropy bipartite alternatives fix expected rather than realised margins (Saracco et al., 2017; Neal, Domagalski and Sagan, 2021; Neal et al., 2024) and are not used.

The counting rule is the methodological point of this section. Swaps are counted by **proposals**, so a rejected proposal is retained as a self-loop of the chain and the stationary distribution is uniform over the fibre of graphs sharing the two degree sequences. Counting only **accepted** swaps instead samples the embedded jump chain and oversamples graphs from which more moves are available. The reference implementation in the legacy deposit advances its loop on accepted swaps and carries this bias, so the numbers of record here come from a proposal-counted sampler and results under both rules are reported side by side (Fosdick et al., 2018).

The nulls of record use attrimotif version 1.1.1 (Shao, 2026) through its *null_test* entry point and its *degree_swap* sampler with a user-supplied statistic function, which allows the open path and the mixing term to be scored through the corrected sampler although the package implements neither count. Four ensembles were drawn at chain lengths of 12, 60, 200 and 500 times the edge count in proposals per sample, each with $S = 500$ samples generated as independent restarts within that ensemble; the four ensembles share the seed 12345 and are not independent of one another. The package default is 12 times the edge count, 7,644 proposals on this network. For every ensemble we report the null mean, its Monte Carlo standard error $sd/\sqrt{S}$, the null standard deviation, the 2.5 % and 97.5 % null quantiles and the z-score, and for the mixing share of the open-path deficit a bootstrap 95 % interval from 2,000 resamples of the ensemble. The null means of the three tested statistics agree across chain lengths to within 0.1 % of their value while the estimated null standard deviations vary between ensembles (Table 10), so the longest chain, 500 times the edge count (318,500 proposals per sample), is taken as the record and the whole series is reported; that choice was made after all four ensembles had been computed and is a reporting convention, not a pre-specified rule; a principled stopping rule exists for sampling bipartite networks with fixed degree sequences (Neal, 2025), it was not applied here, and that is a limitation; the mean acceptance rate is 0.130 at every chain length. A further local proposal-counted re-run of 1,000 samples at 60 and 200 times the edge count of 637 serves as the second implementation. The scored statistics are the four fan classes, an internal negative control that must return a null standard deviation of zero, together with C4, P4 and the mixing term, that is, the degree–degree mixing term Σ over edges of $(d_a - 1)(d_o - 1)$, and, as a descriptive re-expression of that term rather than a fourth tested statistic, the edge degree assortativity coefficient r (Newman, 2002).

On dependence, the mean ± sd describes the null ensemble and is not a confidence interval for a population quantity, and Markov chain samples are not described as independent observations. The text reports z-scores and the null quantiles of Table 7 in preference to empirical p-values; where an empirical p-value is reported, the sample count S and the resolution floor $1/(S + 1)$ are stated with it.

## The exact decomposition of the open path

The decomposition rests on an identity. For a simple bipartite graph, $P4 = \Sigma$ over edges (a,o) of $(d_a - 1)(d_o - 1) - 4 \cdot C4$, where the sum runs over edges and C4 is the number of four-cycles. It is verified unit by unit on the observed graph, where it holds exactly (Results).

Writing M for the mixing term, the identity gives $\Delta P4 = \Delta M - 4\Delta C4$. Applied to the differences between the observed values and the null means, it splits the open-path deficit into two components whose sum must equal that deficit exactly, and this arithmetic self-consistency is displayed in the Results as a check. Reported for each statistic are the raw count, the null mean and standard deviation, the standardised z, and the percentage split $\Delta M/\Delta P4$ against $-4\Delta C4/\Delta P4$, quoted to one decimal place; individual run values appear in Tables 7 and 8.

The decomposition is component accounting at the null means: it is not described as a cause or a mediating pathway, and it identifies no motif-level effect. Because the split is an identity, the percentage carries no sampling uncertainty of its own; uncertainty enters only through the estimated null means, so it is accompanied by a bootstrap interval over the null ensemble and by its range across chain lengths. A point split of this kind can still fail to identify which component carries the departure, and can be unstable under perturbation of the data. Both are checked in the Results, through the joint null distribution of the two components, leave-one-tourist-out re-runs and re-runs at higher rating thresholds, none of which was part of the original analysis plan.

### Descriptive cross-dataset analysis

Within each motif size, class shares and the normalised entropy of the class distribution are computed for both sources. These are descriptive quantities: of them, only the size-3 fan-out to fan-in ratio is degree-determined, since the size-4 denominator also contains the variable four-cycle and open-path counts. The association between the tourism and luxury size-4 profiles is computed across the four size-4 classes and reported as an effect size only, with the number of classes and the exhaustive permutation floor of 0.0417 stated alongside it. No significance test is performed between the datasets, and no domain-level effect size is reported (Shao, Ieiri and Hishiyama, 2024).

### Attribute re-attachment and synthetic controls

Attribute re-attachment is a reproducible workflow and nothing more. The operator $\Phi$ enumerates typed motif instances, attaches to each a summary of its participating edge attributes, and reports the full per-class attribute distribution $D(c)$. No attribute-preserving null exists in the implementation, so $D(c)$ is descriptive and no adjusted inference is drawn from it; $\Phi$ is retained as implementation shorthand and is not described as a framework or as carrying information gain. The operator is exercised on the synthetic panel only: the luxury per-instance records were not accessed and no per-instance attribute is defined on the tourism edge list, so no real-data distribution is compared with or without $\Phi$.

Two synthetic controls act as falsification tests of the pipeline rather than as evidence of external validity. The first plants a single extreme instance value in one of two size-3 classes with similar central summaries, so that the planted maximum stays invisible to count, mean and 99th percentile summaries. The two classes are not matched in the design sense: no tolerance was set, and they differ in instance count by about a factor of four (Results). The second plants four-cycle clustering and scores the four-cycle count against a degree-matched control. The four fan classes are the negative control, whose null standard deviation must be 0. The values are reported in Results.

### Reproducibility, ethics, and data

The nulls and the degree-determinacy guard are called from attrimotif 1.1.1, archived at 10.5281/zenodo.22259239 (Shao, 2026). The predecessor framework package, which carries the corrected proposal-counted sampler and the synthetic validation script, is archived under the concept DOI 10.5281/zenodo.21122972 (MIT licence): the version published before this study (v4, DOI 10.5281/zenodo.21441291; earlier versions were withdrawn by the author) shipped the accepted-count rule in *degree_swap*, and version 5, which carries the proposal-counted sampler, is published under that same concept DOI, with version DOI 10.5281/zenodo.22287908, so no number of record comes from the superseded implementation. That package predates this article and does not contain its analysis. The scripts, cached null ensembles and figure code of this article, together with the software versions, seeds, proposal budgets, sample counts and expected outputs, are archived separately at 10.5281/zenodo.22308052.

The anonymised 17 × 80 rating matrix underlying the tourism network is archived on Zenodo at 10.5281/zenodo.22299150 under CC BY 4.0; the consent obtained in the data-collection study (Ieiri, Nakajima and Hishiyama 2018) covers public re-use of the de-identified matrix, as confirmed by the author with that study's authors, so access does not depend on contacting the author. The luxury quantities are the per-class aggregates already published (Shao, Ieiri and Takahashi, 2025b); the underlying transaction records are proprietary and were not accessed for this study. No new human-subject data were collected: the tourism edge list is an anonymised rating matrix reused from the Kyoto data-collection study (Ieiri, Nakajima and Hishiyama, 2018) through the 2021 analysis of it (Shao, Ieiri and Hishiyama, 2021), and the luxury side uses published aggregates only.

### Use of generative AI

Claude (Opus 5), developed by Anthropic, was used solely to polish the English expression of the manuscript text. It was not used to generate, analyse or interpret any data, results or claims. The author retained the original and revised text, checked that this use complied with the tool's terms and was suitable for the purpose, verified the originality and accuracy of all content and references, and takes full responsibility for the article.

## Results

### Data sources and reconstruction quality

The two sources do not carry the same analytical permissions (Table 5). The tourism network of 17 tourists, 80 sightseeing sites and 637 edges, an edge being a rating greater than zero, reproduces the profile computed in the withdrawn manuscript from the same matrix exactly, 84.7 % fan-out and 15.3 % fan-in, which fixes the edge list analysed here. The node counts, 17 tourists and 80 sites, are the ones stated in the 2021 source study (Shao, Ieiri and Hishiyama, 2021); that study reports no per-class size-3 census, so no published profile is available as a second check. The luxury aggregates are the per-class instance counts published for the 36 monthly customer by item networks of Table 5 (Shao, Ieiri and Takahashi, 2025b), with no per-month node and edge cardinalities in the source and no enumeration here. Every edge-level analysis below is therefore confined to the tourism network.

**Table 5. Availability and descriptive profile of the two data sources.** Cells marked "not reported in the source" were not estimated. The two sources do not carry the same analytical permissions: the tourism network is the inferential worked example, while the luxury column is a provenance-limited illustration whose published item-side counts do not satisfy the minimum edge count the counting identity implies for the transaction total reported alongside them (see below), so no determinacy verdict is drawn from it.

| Item | Tourism network | Luxury aggregates |
|---|---|---|
| Unit analysed | one snapshot, tourists by sites | 36 monthly networks (2019–2021), customers by items |
| Agents / objects / edges (count) | 17 / 80 / 637 | not reported in the source; 26,451 transactions |
| Availability and check | anonymised edge list (T1–T17, no personal identifiers); the size-3 profile of the withdrawn manuscript reproduced exactly | class-level counts only; not reconstructable |
| Size-3 shares (fan-out / fan-in) (%) | 84.7 / 15.3 | 0.175 / 99.825 |
| Size-4 shares (fan-out / four-cycle / open path / fan-in) (%) | 65.28 / 8.48 / 24.37 / 1.87 | 0.004 / 0.015 / 0.578 / 99.403 |
| Class counts 3-A / 3-B / 4-A / 4-B / 4-C / 4-D (motif instances) | exact census, this study | 3,910 / $2.23\times10^{6}$ / 1,560 / 5,350 / $2.11\times10^{5}$ / $3.63\times10^{7}$ |
| Normalised entropy (size-3 / size-4) | 0.617 / 0.654 | 0.019 / 0.027 |
| Edge-level null feasible | yes | no (no edge list) |

Supports: edge-level analysis for tourism, class-level description only for luxury; the reconstruction is exact. Does not support: any node or edge count for the luxury networks, or a symmetric comparison.

## Fan motif counts are degree-determined

Under a null that fixes both degree sequences, the four fan classes have no variance to test against. On the synthetic control the enumerated fan counts are 720 / 1200 / 480 / 1600 and every null sample reproduces each of them, giving a null standard deviation of 0. The same follows from the star identity (Wegner, 2014), and it holds on the tourism network, where the prediction $\Sigma_v C(d_v, k)$ and the exact enumeration agree unit for unit. A significance statement about a fan class under this null is therefore not a weak result; it is undefined by construction (Table 6, Fig. 1).

**Table 6. Motif class registry and degrees of freedom under a degree-preserving null.** The identity is Wegner's (Wegner, 2014); it is cited, not claimed.

| Class | Composition (agents, objects) | Induced count | Degree-determined | Null d.f. |
|---|---|---|---|---|
| 3-A fan-out | 1, 2 | Σ over agents of C(d, 2) | yes | 0 |
| 3-B fan-in | 2, 1 | Σ over objects of C(d, 2) | yes | 0 |
| 4-A fan-out | 1, 3 | Σ over agents of C(d, 3) | yes | 0 |
| 4-B overlap (C4) | 2, 2; 4 edges | not a function of the degree sequences | no | non-degenerate |
| 4-C open path (P4) | 2, 2; 3 edges | not a function of the degree sequences | no | non-degenerate |
| 4-D fan-in | 3, 1 | Σ over objects of C(d, 3) | yes | 0 |

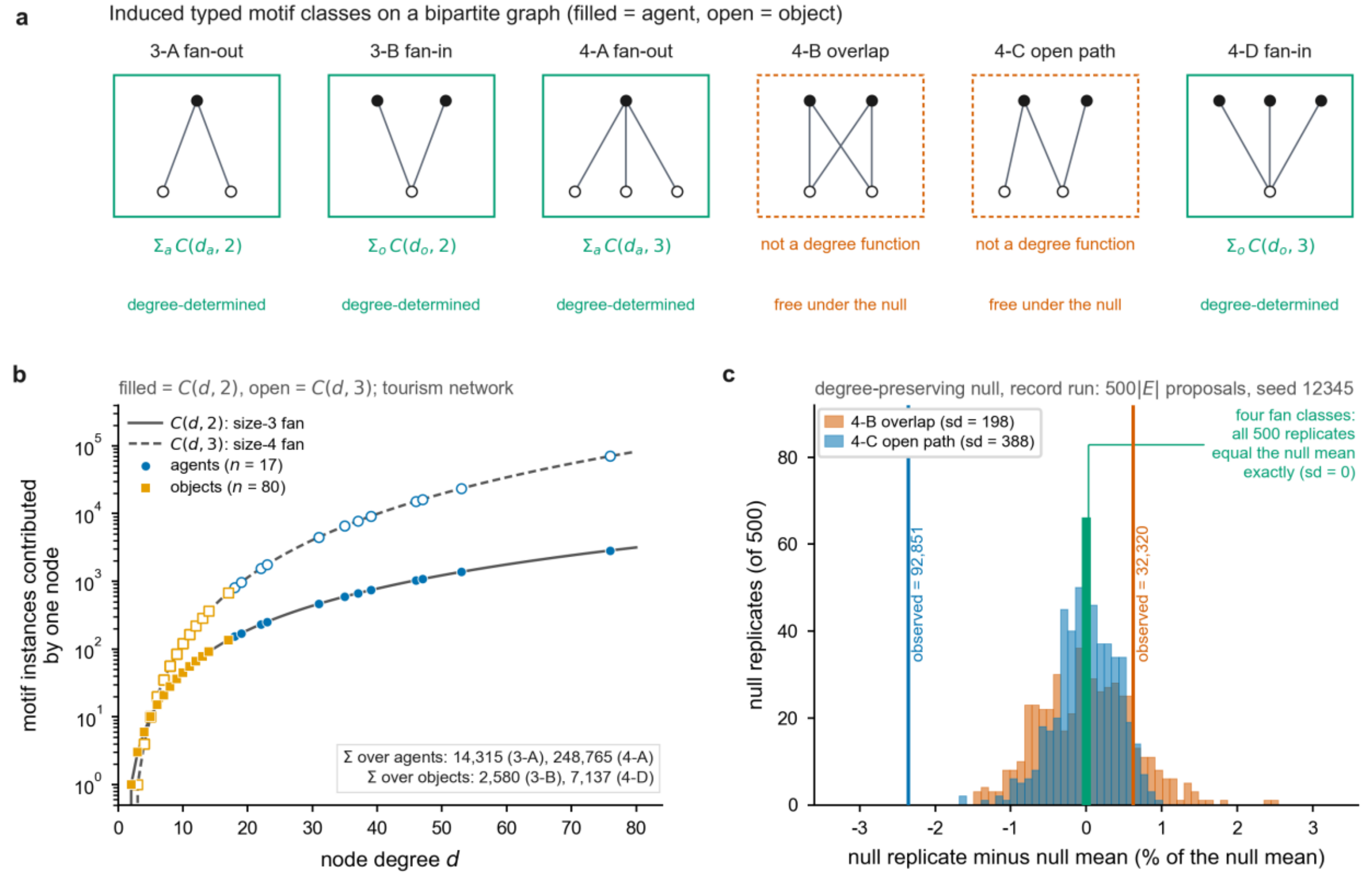

**Figure 1. Fan counts have no degrees of freedom under a degree-preserving null.** (a) Schematics of the six typed classes, the counting formula on each fan, the four-cycle and open path marked not degree-determined; no axes. (b) x = node degree d (dimensionless), y = C(d, 2) and C(d, 3) contributions in motif instances; unit of analysis = node. (c) Tourism null ensembles from the record run only, S = 500 at a chain length of 500 times the edge count: x = percentage deviation from the null mean, y = frequency over null samples; the fan classes collapse to one line at zero deviation while the four-cycle and open-path ensembles are dispersed, observed values as vertical rules. Error

representation: the null histogram itself, not a confidence interval. Takeaway: under this null a fan count cannot move, and only the non-fan statistics have a distribution to test against.

Supports: fan-class significance under this null is vacuous; the degree sequences recover the counts exactly. Does not support: a new theorem, or the claim that all motif classes are degree-determined.

### The cross-dataset profile contrast restated

The contrast between the profiles is large, and on the tourism side part of it is restated by the degree sequences. In the tourism network the raw fan counts are exact functions of the two degree sequences, and so is the size-3 fan-out to fan-in ratio, whose two components are both fans: size-3 fan-out is 84.7 %, and that ratio states which side of the network holds the concentrated degree sequence. The luxury figure set beside it, 0.2 % fan-out or equivalently 99.8 % fan-in, is a published descriptive value and nothing more here: the counts it is computed from do not satisfy the edge bound the same identity implies (next subsection), so this note draws no determinacy verdict on that side. Within-size shares at size 4 are not degree-determined, because their denominator also contains the four-cycle and open-path counts, which vary across the fixed-degree ensemble; the size-4 fan-in share of 1.9 % against 99.4 %, and the normalised entropies that separate the two profiles in the same direction (Table 5), are therefore reported as descriptions and not as degree-sequence restatements. The within-size-4 association is descriptive and one-sided as computed: Pearson $r = -0.543$ with one-sided exhaustive permutation $p = 0.125$ and two-sided $p = 0.375$; Spearman rho = $-0.80$ with one-sided $p = 0.167$ and two-sided $p = 0.333$; $n = 4$ with a permutation floor of $1/24 = 0.0417$. The permutation runs over four class labels rather than over any sampling unit, so these are re-orderings of a four-point description; at that floor no p value here can reach a conventional threshold, and none is offered as a test.

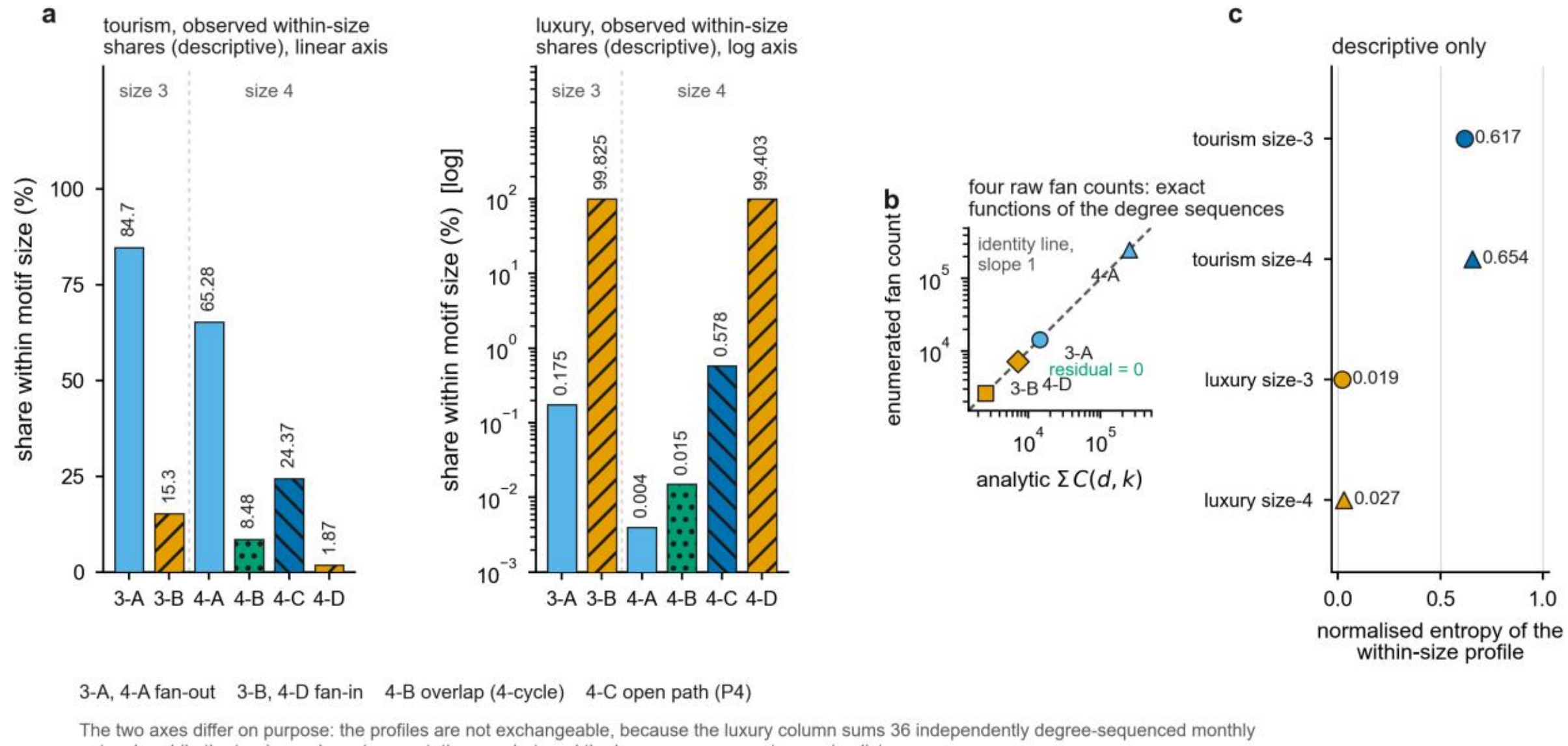

**Figure 2. Raw fan counts are reproduced by the degree sequences; within-size shares at size 4 are not.** (a) Faceted bars: x = motif class (two size-3, four size-4), y = observed within-size share in per cent, faceted by dataset on separate axes so no shared total is implied, values from Table 5 printed on the bars; unit of analysis = motif class within a dataset. This panel is a descriptive comparison of observed shares and carries no determinacy claim, since the size-4 denominators contain the variable four-cycle and open-path counts. (b) Tourism only: x = analytic prediction Σ C(d, k), y = observed fan count, both in motif instances; the four raw fan counts lie on the identity line with no residual, which is the panel that demonstrates exact determinacy. (c) Normalised entropy, four points, labelled descriptive. Error representation: none, because both censuses are exhaustive enumerations rather than samples; the luxury side has no edge list and therefore no panel (b). Takeaway: in the tourism network the raw fan counts, and the size-3 fan-out to fan-in ratio, are reproduced by the two degree sequences, while the size-4 within-size shares are not.

Supports: in the tourism network the raw fan counts and the size-3 ratio are a degree-sequence signature; the association is an effect size, not a test. Does not support: the same reading for size-4 within-size shares, the same verdict on the luxury aggregates, a contrast of structural roles, or any domain-level regularity.

**The published luxury aggregates do not satisfy the edge bound the identity implies.** The same identity that fixes the fan counts also bounds the edge count from below. Because 3-B is the sum over objects of C(d, 2) and 4-D the sum over objects of C(d, 3), the two published item-side counts, $2.23\times10^6$ and $3.63\times10^7$, jointly require at least 89,502 customer-item edges across the 36 monthly networks, a minimum attained by a two-degree relaxation at object degrees of about 50 to 51, whereas 26,451 transactions are reported for the same networks (Shao, Ieiri and Takahashi, 2025b). This note cannot reconcile the two from published figures. Multi-edges between one customer and one item, pooling that counts an item across months, or an edge definition in the source that differs from the one used here would each account for the gap, and no edge list is available to decide between them. The luxury side is therefore a provenance-limited illustration of the counting identity and carries no determinacy verdict of its own; the tourism matrix is the sole inferential worked example in this note, and the tourism counts pass the same bound on both sides.

### Departures from the degree-preserving null on the tourism network

Only the statistics that are not degree-determined can depart. Two do: the open path and the mixing term; the four-cycle does not (Table 7, Fig. 3). Against the proposal-counted null ensemble produced with attrimotif 1.1.1 (Shao, 2026) ($S = 500$, seed 12345, mean acceptance 0.130), the four-cycle count is degree-consistent at $z = 1.0$, while the open path departs from its null mean by −2,243 instances, which is 2.4 % of that mean, at $z = -5.8$, and the mixing term at $z = -3.3$; the observed edge degree assortativity is −0.2570, and because it is an affine re-expression of the mixing term within one ensemble it carries that same z by construction rather than a z of its own. Throughout the prose, z is quoted to one decimal place, the precision the Monte Carlo replication supports; the tables and the figure captions carry the computed values, and the sampler-sensitivity comparison below keeps two decimals because the differences it discusses are themselves at the second decimal and are reported there as Monte Carlo variation rather than as resolved differences. The open-path empirical two-sided p is 0.002, computed as $(r + 1)/(S + 1)$ at the floor $1/(S + 1)$ for $S = 500$ samples, so it reads as "no null sample was more extreme" rather than as a calibrated tail probability, and no stopping rule certified the chain length that produced the ensemble behind it (Limitation 6). The three statistics are not three independent readings: the identity $P4 = M - 4 \cdot C4$ holds on every graph in the ensemble, so the family has rank two, and any multiplicity correction applied at family size three, Holm included, is conservative here. Applied at family size three to the record empirical p values, Holm gives adjusted values of 0.0060 for the open path, 0.0120 for the mixing term and 0.311 for the four-cycle count, so no reading changes under the correction. The four-cycle result is a marginal one; the joint position of the four-cycle count and the mixing term relative to the null cloud is reported below and does not resolve which of the two carries the open-path deficit.

**Table 7. Tourism null results, numbers of record (chain length 500 × |E| proposals per sample, S = 500, seed 12345).** Null mean ± sd describes the null ensemble, not a population interval; the Monte Carlo standard error (MCSE) of the null mean and the 2.5 % and 97.5 % null quantiles are given so that the observed value can be placed in the ensemble directly.

| Statistic | Observed (instances) | Null mean ± sd (instances; MCSE in instances) | Null 2.5 % / 97.5 % (instances) | z (unitless) | Contribution to the P4 deficit (instances) |
|---|---|---|---|---|---|
| Overlap (C4) | 32,320 | 32,120.50 ± 198.4 (8.9) | 31,757 / 32,528 | +1.01 | four-cycle term −4 (C4 − null) = −798.0 |
| Open path (P4) | 92,851 | 95,094.35 ± 388.5 (17.4) | 94,241 / 95,763 | −5.77; p = 0.002, S = 500, floor 1/(S + 1) | total deficit −2,243.3 |
| Degree-mixing term Σ (d_a − 1)(d_o − 1) | 222,131 | 223,576.33 ± 434.1 (19.4) | 222,781 / 224,390 | −3.33 | mixing term −1,445.3, that is 64.4 % of the deficit (bootstrap 95 % interval 62.1 to 67.1 %) |
| Assortativity r (unitless; descriptive re-expression of the mixing term, not a tested statistic) | −0.2570 | not applicable; r is affine in the mixing term at fixed degree sequences | not reported | −3.33, equal to the mixing term's z by construction (affine re-expression) | not applicable |
| Four fan classes | exact census | identical to observed, sd = 0 | identical to observed | undefined | not applicable |

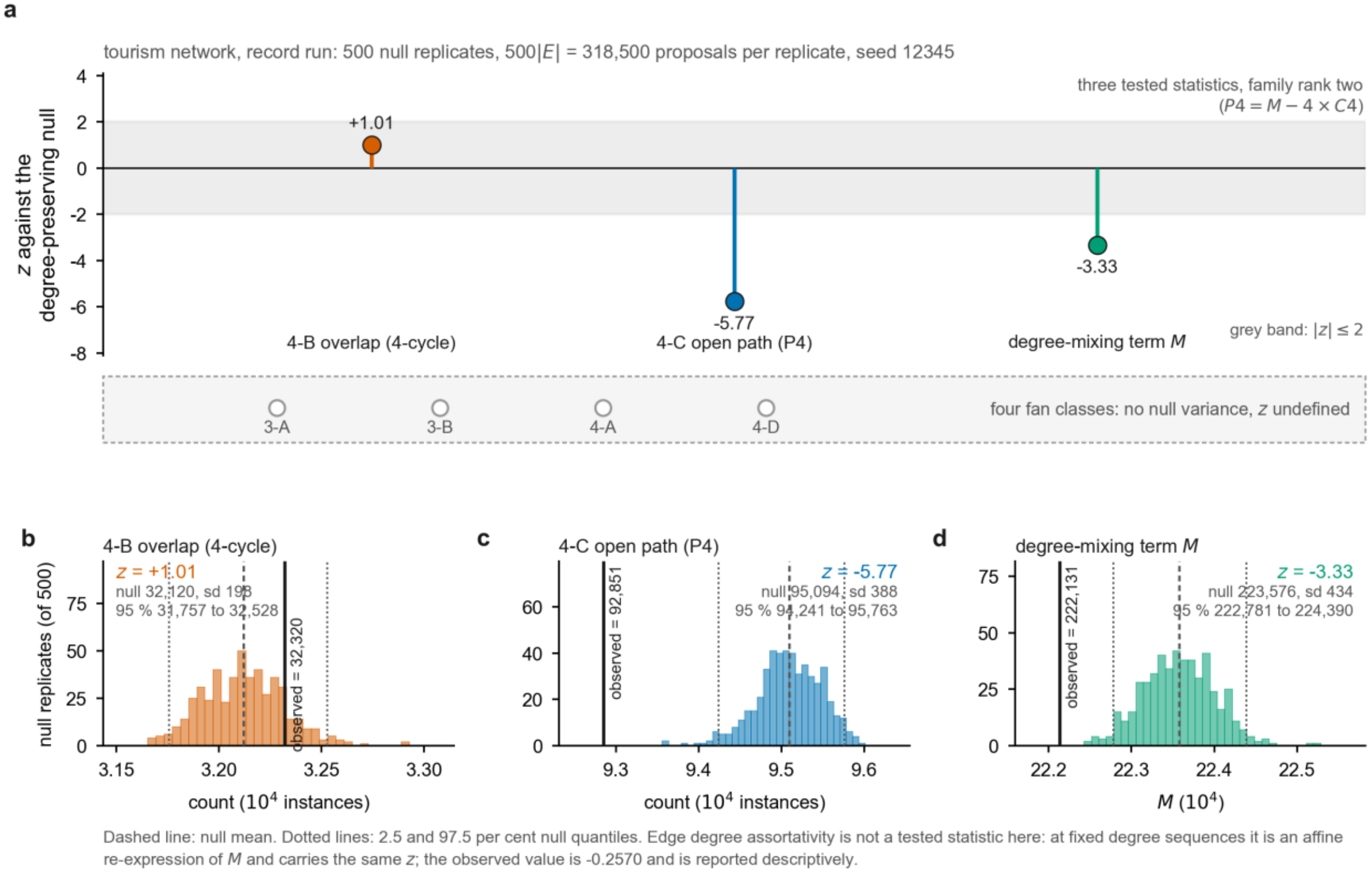

**Figure 3. What departs from the null on the tourism network.** (a) x = the three tested statistics (four-cycle C4, open path P4, mixing term), y = z in null standard deviations; the four fan classes are drawn in a separate strip below the axis, labelled "no null variance, z undefined", rather than placed on the z axis. (b) to (d) one panel per tested statistic, x = the count, y = the number of null replicates, with the observed value as a vertical rule. Unit of analysis = the whole network. Error representation: the histogram of the S = 500 null replicates in each of panels (b) to (d), with the null mean drawn as a dashed line and the 2.5th and 97.5th null percentiles as dotted lines; these describe the null distribution and are not a population confidence interval. The three statistics span two dimensions rather than three, because P4 = M − 4·C4 holds exactly on every graph of the ensemble; the panel labels state this so that the three bars are not read as three independent tests. Takeaway: of the three tested statistics only the open path and the mixing term depart marginally, while the four-cycle is marginally degree-consistent.

Supports: the four-cycle is marginally null-consistent and the open path is not, the departure being a deficit. Does not support: an absence of clustering as an existence claim, a conditional reading of the four-cycle result, since the null ensemble supplies no conditional distribution at the observed mixing term, or any causal reading.

## Exact decomposition of the open-path deficit

The open-path count is not independent of the other two statistics. On any simple bipartite graph P4 equals the sum over edges of (d_a − 1)(d_o − 1) minus four times the four-cycle count, and on the observed network this holds unit for unit: 222,131 − 4 × 32,320 = 92,851. The deficit against the null mean therefore decomposes term by term with no residual. Of the total −2,243.3, the mixing term contributes −1,445.3 and the four-cycle term −4·ΔC4 contributes −798.0, that is 64.4 % and 35.6 % of the deficit respectively (Table 7), the mixing component ranging from 62.7 to 66.9 % across chain lengths. The two percentages carry no sampling uncertainty of their own and inherit only the estimation error of the null means. This is a point accounting at the null means and nothing more. The next subsection shows that it does not identify which component carries the deficit, and that it does not survive leave-one-tourist-out deletion or a change in the rating threshold that defines an edge.

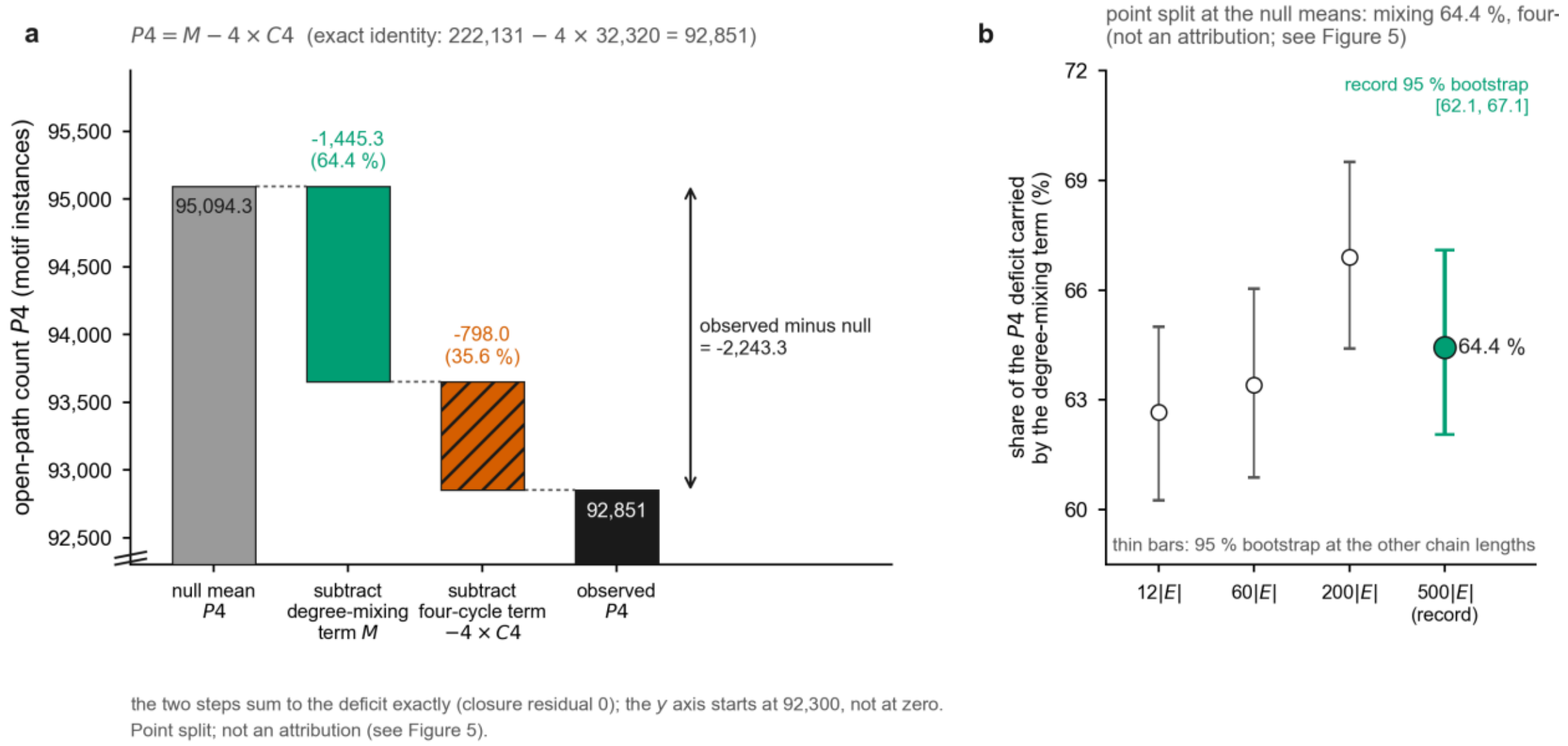

**Figure 4. The open-path deficit decomposes into a mixing term and a four-cycle term.** (a) Waterfall chart: x = decomposition components, y = contribution to the observed-minus-null open-path deficit in motif instances; the waterfall starts at the null mean 95,094.35 and subtracts the mixing term, −1,445.3 (64.4 %), and then the four-cycle term −4·ΔC4, −798.0 (35.6 %), descending to the observed 92,851, with the identity and the sum check to −2,243.3 printed alongside. (b) The mixing share: the record value 64.4 % with its bootstrap 95 % interval, 62.1 to 67.1 %, and the four chain-length points at 12, 60, 200 and 500 times the edge count, 62.7, 63.4, 66.9 and 64.4 %, each with the bootstrap interval listed for it in Table 10; there is no block-split estimate and no separate 60 and 200 times the edge count side panel. Unit of analysis = the whole network. Error representation: (a) endpoints derive from null ensemble means whose standard deviations are in Table 7, and given those means the shares are exact; (b) bootstrap 95 % intervals of the mixing share at all four chain lengths, with the record (500 × |E|) interval emphasised. Takeaway: at the null means the point accounting assigns 64.4 % of the deficit to the mixing term and 35.6 % to the four-cycle term; Fig. 5 shows that this assignment is neither identifiable nor stable.

Supports: an exact point accounting of the deficit into a mixing component and a four-cycle component. Does not support: reading that accounting as an attribution of the deficit to either component, a structural effect residing in the open-path class, or generalisation past this network.

## Attribution is not identifiable and not stable

Three further analyses, all computable from the deposited matrix and the cached ensembles, show that the point accounting above is not an attribution and does not survive perturbation of the data. The joint-null analysis uses the record ensemble, chain length 500 times the edge count with S = 500. The deletion and threshold re-runs use a chain length of 60 times the edge count with S = 500 and seed 12345 throughout; at those settings the full network gives an open-path z of −6.1 and a mixing share of 63.4 %, so the rows below are read against those values and not against the record run.

**The two components are almost collinear under the null, and the observed pair lies outside the null cloud.** Across the record ensemble the mixing term and the four-cycle count correlate at 0.968, so the null cloud lies close to a line and the two coordinates cannot be moved independently within it. Marginally, the observed mixing term lies below all 500 samples drawn (S = 500): the smallest sampled value is 222,427 against an observed 222,131, so 0 of 500 samples sit at or below the observed value. The observed four-cycle count of 32,320 is by contrast not low: a fraction 0.140 of the null samples sit at or above it. Because the ensemble provides no sample at the observed mixing level, the null supplies no conditional distribution there, and any expectation of the four-cycle count conditional on the observed mixing term would be an extrapolation beyond the sampled support; none is reported here. What the joint null does support is a distance: the observed pair lies far outside the null cloud, at a Mahalanobis distance of 17.3, which is reported as a distance and not as a test, since no distributional form is assumed for the joint null. Neither the marginal split nor the joint position identifies which of the two components carries the departure. The point split is an accounting of two numbers, not an identification of which one carries it.

**The split is unstable under leave-one-tourist-out deletion.** Deleting one tourist at a time and re-running the null, the leave-one-tourist-out re-analysis, gives 17 re-analyses, of which 16 are distinct, because T16 and T17 have identical binarised rating rows (Table 8). The open-path deficit survives every one of them, with z between −4.5 and −8.1. The mixing share does not: it runs from 36.6 to 116.5 %, with a median of 65.2 %, and a share above 100 % means the four-cycle term moves against the deficit rather than with it. The two most active tourists rated 76 of the 80 sites each; dropping both leaves 485 edges, and on that subgraph the mixing share falls to 15.1 %, the mixing z to −1.2 and the four-cycle z rises to +3.8 while the open-path z is −9.3. There the deficit is carried by the four-cycle term, which is the first of the failure conditions this article states in the Discussion; the re-analysis triggers it.

**The deficit itself depends on the edge threshold.** The edge rule of record places an edge where the rating is greater than zero. Raising that threshold (Table 9) leaves the deficit in place at rating > 1, z = −5.0, and removes it above that: z = −2.5 at rating > 2, −1.6 at rating > 3 and −1.4 at rating > 4, on networks of 547, 331 and 135 edges. The mixing share is reported along that series only where the open-path deficit is distinguishable from zero, that is where |z| is at least 1.96; where it is defined it is not stable either, and at rating > 3 and rating > 4 it is undefined, because the deficit that forms its denominator is not distinguishable from zero and a ratio taken there carries no information about the stability of the split. Part of this is loss of power, since the graph shrinks as the threshold rises, and part of it is a change in the object analysed, since nodes that keep no edge leave the analysed graph. The two cannot be separated on one dataset, so the statement of record is that the deficit is a property of the rating > 0 graph and is not shown to hold under other edge definitions.

Taken together, the deficit is extreme relative to the sampled hard-degree null for this observed graph, the sense in which it is called a departure throughout, and it is robust to which single tourist is dropped, while its class-level interpretation is undetermined and unstable. We therefore report the deficit and withhold the attribution.

**Table 8. Leave-one-tourist-out re-analysis, and the two most active tourists removed together.** Every row is an independent re-run at a chain length of 60 times the edge count with S = 500 samples and seed 12345, on the graph remaining after the deletion. Sites rated is the number of sites the dropped tourist rated above zero. Mixing share is the mixing component of the open-path deficit, so a value above 100 % means the four-cycle component works against the deficit. T16 and T17 have identical binarised rating rows, so their two rows coincide. The full network at these settings gives z P4 = −6.14 and a share of 63.4 %.

| Tourist dropped | Sites rated by that tourist | Edges remaining | z C4 | z P4 | z M | Mixing share (%) |
|---|---|---|---|---|---|---|
| T1 | 35 | 602 | +1.66 | −6.56 | −2.67 | 47.3 |
| T2 | 35 | 602 | +1.61 | −6.48 | −2.74 | 48.7 |
| T3 | 39 | 598 | +0.29 | −5.07 | −3.67 | 87.9 |
| T4 | 53 | 584 | +1.67 | −7.27 | −2.90 | 50.2 |
| T5 | 37 | 600 | +0.89 | −6.19 | −3.59 | 69.7 |
| T6 | 19 | 618 | +1.14 | −6.16 | −3.35 | 61.7 |
| T7 | 47 | 590 | +0.75 | −5.52 | −3.17 | 71.0 |
| T8 | 39 | 598 | +0.90 | −5.91 | −3.35 | 67.9 |
| T9 | 23 | 614 | +1.37 | −5.97 | −2.66 | 51.9 |
| T10 | 76 | 561 | +2.40 | −8.11 | −2.44 | 36.6 |
| T11 | 76 | 561 | +2.26 | −7.66 | −2.38 | 37.5 |
| T12 | 46 | 591 | +0.56 | −5.97 | −3.75 | 79.5 |
| T13 | 10 | 627 | −0.41 | −5.09 | −5.22 | 116.5 |
| T14 | 18 | 619 | +0.33 | −5.45 | −4.21 | 87.7 |
| T15 | 22 | 615 | +1.06 | −5.04 | −2.60 | 57.1 |
| T16 | 31 | 606 | +0.77 | −4.45 | −2.67 | 65.2 |
| T17 | 31 | 606 | +0.77 | −4.45 | −2.67 | 65.2 |
| T10 and T11 removed together | 76 each | 485 | +3.77 | −9.26 | −1.20 | 15.1 |

**Table 9. Sensitivity of the tourism results to the rating threshold that defines an edge.** Each row is an independent re-run at a chain length of 60 times the edge count with S = 500 samples and seed 12345. The first row is the edge rule of record. Node counts are the tourists and sites that keep at least one edge under the rule; nodes that keep none contribute zero to every count and leave the analysed graph. The mixing share is the mixing component of the open-path deficit, so it is defined only where that deficit is distinguishable from zero; it is reported where |z P4| is at least 1.96 and marked undefined at the two highest thresholds, where the denominator is not distinguishable from zero and the ratio carries no information about the split.

| Edge rule | Tourists / sites with at least one edge | Edges | z C4 | z P4 | z M | Mixing share (%) |
|---|---|---|---|---|---|---|
| rating > 0 (record) | 17 / 80 | 637 | +1.07 | −6.14 | −3.26 | 63.4 |
| rating > 1 | 17 / 80 | 621 | +0.62 | −5.00 | −3.11 | 74.2 |
| rating > 2 | 17 / 79 | 547 | +0.39 | −2.52 | −1.26 | 65.6 |
| rating > 3 | 17 / 76 | 331 | −0.85 | −1.64 | −1.77 | undefined (deficit not distinguishab le from zero) |
| rating > 4 | 16 / 62 | 135 | −0.63 | −1.43 | −1.21 | undefined (deficit not distinguishab le from zero) |

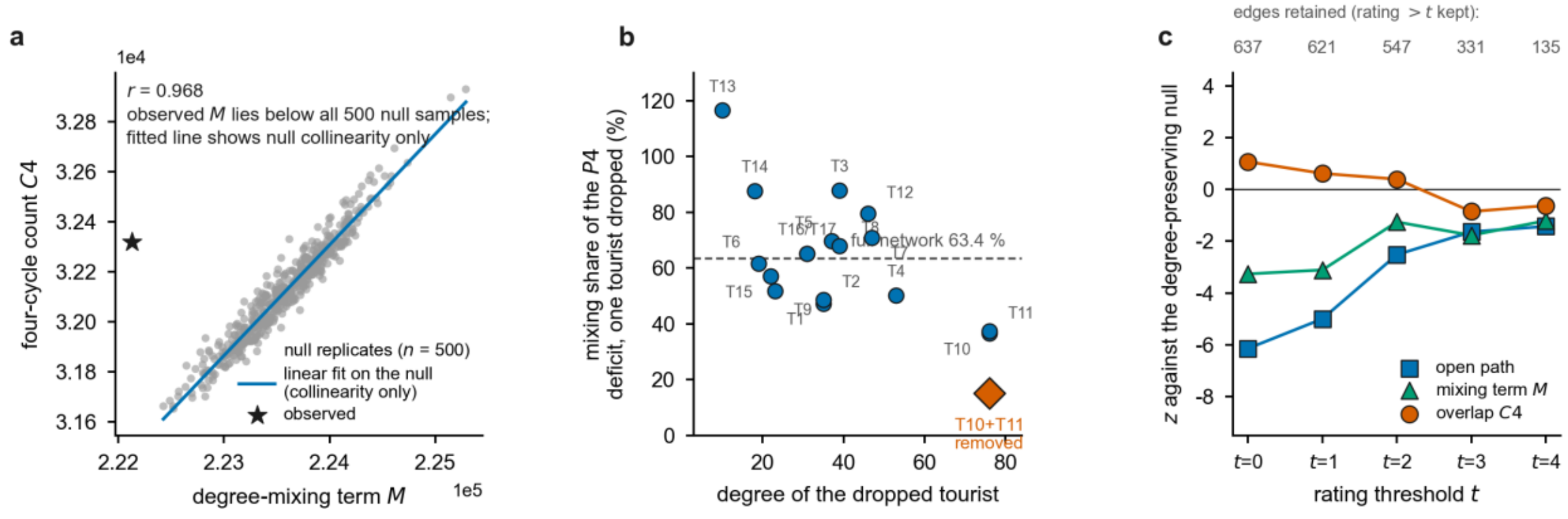

**Figure 5. Attribution instability.** (a) Null scatter of the mixing term against the four-cycle count on the record ensemble, with the observed pair outside the sampled range of M; the fitted line is shown only to display the null collinearity, and no conditional expectation is read off it at the observed value. (b) Mixing share under leave-one-tourist-out against the dropped tourist's degree, with the drop-two point; (c) open-path, mixing and four-cycle z against the rating threshold. Unit of analysis = the whole network under one perturbation. Panel (a) uses the record ensemble, S = 500, and marks the null correlation 0.968 together with the fact that the observed mixing term lies below all 500 samples drawn; panels (b) and (c) use the re-runs of Table 8 and Table 9. Error representation: (a) the null ensemble itself; (b) and (c) points, one per re-run, with no interval, because each point is a separate exhaustive enumeration scored against its own null. Takeaway: the point split is not identified by the data and does not hold under leave-one-tourist-out deletion or a change in the edge rule.

Supports: the open-path deficit as a finding about the rating > 0 graph under this null, and the instability of its decomposition as a finding in its own right. Does not support: an attribution of the deficit to degree mixing rather than to the four-cycle term, the opposite attribution, or the transfer of either to another edge rule.

### Sampler sensitivity and the correction disclosure

The counting rule of the edge-swap sampler (Methods) was found during this reanalysis to change the reported numbers, and we state it rather than absorb it. The record ensemble uses the longest of the four chain lengths described in Methods, 500 times the edge count in proposals, that is 318,500 per sample. Across those chain lengths the null mean of each tested statistic moves by less than 0.1 % of its value: the open-path null mean runs from 95,058.99 to 95,102.49, a spread of 43.5 instances against Monte Carlo standard errors of 16.1 to 17.4 per ensemble. What moves is the estimated null standard deviation, 359.6, 366.6, 361.8 and 388.5 respectively, itself varying non-monotonically, and it is that variation rather than any drift in the null mean that carries the open-path z from −6.25 to −5.77. A bootstrap of the null sd itself (5,000 resamples of each ensemble) gives 95 % intervals of 335.8 to 382.4, 343.2 to 388.8, 336.7 to 385.8 and 361.0 to 414.7, which overlap, so the drift in z is within the sampling uncertainty of the sd estimates rather than evidence of a trend. A second ensemble at the record chain length with a different seed (54321) gives an open-path z of −6.44 with null sd 346.1, a four-cycle z of 1.06, a mixing-term z of −3.50 and a mixing share of 65.2 % (bootstrap 95 % interval 62.8 to 67.6 %). The seed-to-seed difference in the open-path z at one chain length, 0.66 computed from the unrounded z values (rounded values give 0.67), exceeds the drift across chain lengths, which is consistent with Monte Carlo variation of the estimated sd rather than with a trend; two seeds cannot establish more than that, and every reading is unchanged. The record ensemble was chosen after all four chain lengths had been computed, as a reporting convention rather than a pre-specified rule, and its open-path z, −5.77, is the least extreme of the four; the reported sampler shift is therefore measured against the most conservative of the available ensembles. No stopping rule certified any of these chain lengths: a principled criterion for randomly sampling bipartite networks with fixed degree sequences was available (Neal, 2025) and was not applied here, so no ensemble reported above is certified as sufficient by a pre-specified rule, and the calibration of every z and every quantile in this article rests on that uncertified chain length (Limitation 6). Every reading is the same at every chain length, and no plateau is claimed from the trace; the longest chain is taken as the record and the whole series is disclosed (Table 10, Fig. 6). Over the same series the four-cycle z ranges from 0.98 to 1.11 and the mixing z from −3.55 to −3.26. The mixing share of the deficit ranges from 62.7 % to 66.9 %, with bootstrap 95 % intervals that together span about 60 % to 70 %; the record value is 64.4 % (62.1 to 67.1 %). That is the variation attributable to the sampler alone; the much wider variation of the same share under leave-one-tourist-out deletion and under the edge rule is reported above and is not a sampler effect. Inside the sampler-correction disclosure only, that is, this paragraph, Table 10, Fig. 6, the sub-question D row of the research-question matrix and Limitation 6: the superseded accepted-count rule, which advanced its chain to 15 times the edge count in accepted swaps, put the mixing share at 66.2 %. The shift quoted here is defined as the change in z from that legacy rule to the four proposal-counted chain-length ensembles of Table 10, and its largest value is 0.27 on the open path, whether measured against the record run or across all four chain lengths. Two wider comparisons are larger and are stated as such: against the older local proposal-counted re-runs the largest shift is 0.31 on the mixing term at 200 times the edge count, where the legacy −3.45 becomes −3.14; against the corrected deposit script it reaches 0.35 on the open path, where the legacy −6.04 becomes −6.39, a comparison that changes the script and the sample count as well as the counting rule. No reading changes under any of them. The deposit version published before this study (v4) shipped the accepted-count rule; version 5, which carries the proposal-counted sampler, is published under that same concept DOI, with version DOI 10.5281/zenodo.22287908; the numbers of record come from attrimotif 1.1.1.

**Table 10. Sensitivity to chain length, counting rule and implementation.** Each chain-length row is an ensemble of S = 500 samples drawn with the attrimotif 1.1.1 proposal-counted sampler as independent restarts within that ensemble; the four ensembles share the seed 12345 and are not independent of one another. The P4 null mean column carries its Monte Carlo standard error in parentheses, and the share column the bootstrap 95 % interval (2,000 resamples). The null means agree across chain lengths to within 0.1 % of their value while the estimated null standard deviations vary between ensembles, which is what moves the open-path z. The accepted-count row is superseded and supports no conclusion.

| Setting | C4 z | P4 null mean (instances; MCSE in instances) | P4 z | Mixing z | Mixing share (%) [95 % interval] | Acceptance (proportion) |
|---|---|---|---|---|---|---|
| Chain length 12 × \|E\| (attrimotif default) | +1.11 | 95,099.63 (16.1) | −6.25 | −3.32 | 62.7 % (60.3 to 65.0) | 0.130 |
| Chain length 60 × \|E\| | +1.07 | 95,102.49 (16.4) | −6.14 | −3.26 | 63.4 % (60.9 to 66.0) | 0.130 |
| Chain length 200 × \|E\| | +0.98 | 95,058.99 (16.2) | −6.10 | −3.55 | 66.9 % (64.4 to 69.5) | 0.130 |
| Chain length 500 × \|E\| (record) | +1.01 | 95,094.35 (17.4) | −5.77 | −3.33 | 64.4 % (62.1 to 67.1) | 0.130 |
| Accepted-count sampler (legacy deposit through v4; 15 × \|E\| accepted swaps), superseded | +0.99 | not applicable | −6.04 | −3.45 | 66.2 % | not applicable |

Two further proposal-counted implementations reproduce the series within Monte Carlo noise and are not part of the shift definition: a local script with S = 1,000 samples at 60 and 200 times the edge count gives four-cycle z of +1.11 and +1.03, open-path z of −6.21 and −6.00, mixing z of −3.23 and −3.14 and shares of 62.2 % and 63.3 %; the corrected deposit script (version 5, 200 times the edge count; C4 with S = 300, P4 with S = 500) gives four-cycle z of +1.03 and open-path z of −6.39. In that local re-run the edge degree assortativity coefficient itself scores z = −3.23 at 60 times the edge count, equal to the mixing-term z of the same ensemble, as the affine relation between the two requires.

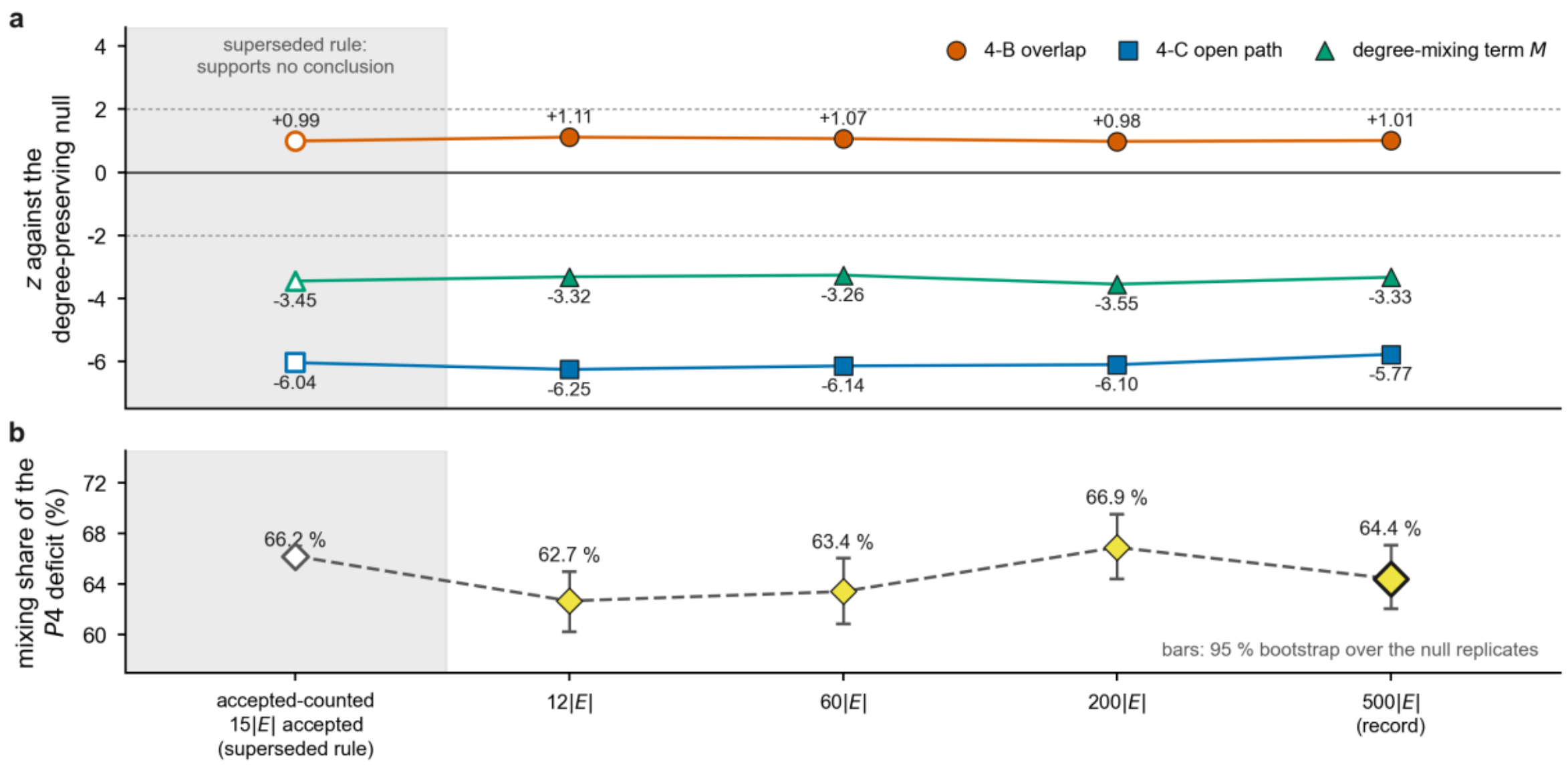

Null means move by less than 0.1 % across chain lengths; the open-path *z* drifts from -6.25 to -5.77 because the estimated null sd varies (359.6 / 366.6 / 361.8 / 388.5); the reading is the same at every chain length.
Largest absolute *z* shift from the superseded accepted-count rule: 0.27 (open path, record run). Against the older local re-runs it is 0.31 (mixing term), and against the corrected deposit script 0.35 (open path).
Open markers mark the superseded rule; the enlarged diamond marks the record run.

**Figure 6. The swap-counting rule matters for reporting, not for the conclusion.** In both panels x = chain length in multiples of the edge count (12, 60, 200, 500; the record at 500) with the superseded accepted-count point marked. (a) y = z for the four-cycle, open-path and mixing statistics as three lines. (b) y = mixing share of the deficit in per cent with its bootstrap 95 % interval. Unit of analysis = the whole network under one sampler configuration. Error representation: points; Monte Carlo standard errors of the null means are reported in Table 10, and no error bar is drawn on z. The accepted-count points are annotated as superseded. Takeaway: the null means agree across chain lengths to within 0.1 % of their value while the estimated null standard deviations vary between ensembles, so the open-path z drifts from −6.25 to −5.77 with no reading changing; the share point estimates range from 62.7 % to 66.9 %, and the largest z shift from the superseded rule to those four ensembles is 0.27, which is why the rule and the chain length must still be stated.

Supports: the readings hold across budgets and implementations; the rule and deposit state must be disclosed. Does not support: further use of the superseded accepted-count numbers as evidence.

## Synthetic controls and attribute re-attachment

The pipeline detects planted structure when it is present. With a planted four-cycle cluster the four-cycle statistic gives z = 239.4 against z = −0.8 for the degree-matched control; these values of record are produced by version 5 of the validation script, which is now the copy carried both in the analysis code and in version 5 of the deposit (10.5281/zenodo.22287908), and an independent re-run of the same design with the attrimotif 1.1.1 sampler gives +239.0 and −0.9, so the degree-matched control satisfies $|z| < 1$ under both. Attribute re-attachment behaves the same way on synthetic data: two size-3 classes with similar central summaries, mean 0.33 against 0.30 and 99th percentile 0.36 against 0.35, differ at the extreme, planted maximum 6.48 against 0.38, a ratio of 16.9, which the summaries do not reveal and the full per-class distribution does. The two classes are not matched: no tolerance was set on the summaries, and they hold 2,838 and 11,710 instances respectively, so the comparison shows only that equal-looking central summaries can hide an extreme value. On the real side no comparison of that kind is defined: the luxury data enter only as published aggregates, whose per-class mean returns span 0.25 to 0.32 with a single-instance maximum of 6.24, while the per-instance records were not accessed. Attribute re-attachment is therefore demonstrated on the synthetic panel only; the absence of a real-data gain is reported as such and is not evidence that no gain exists.

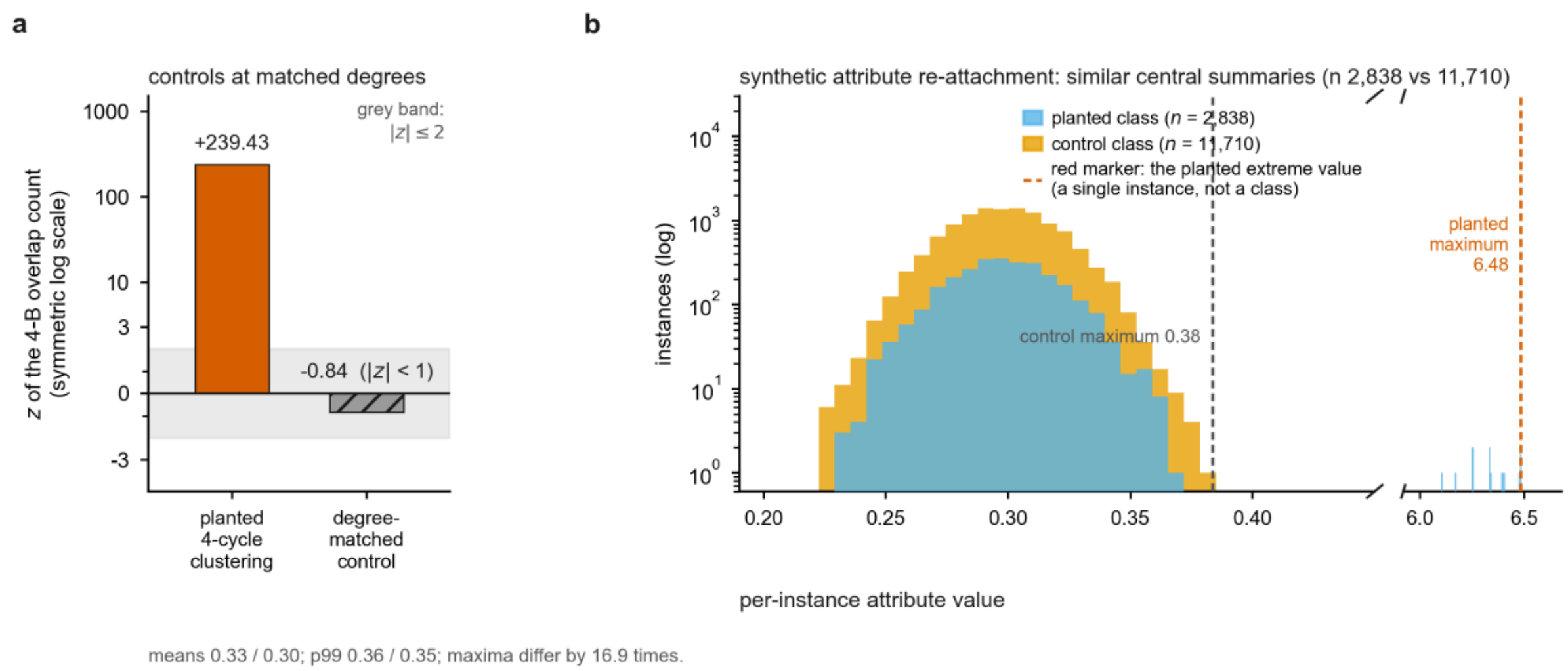

**Figure 7. Controls for the pipeline, and the null result for attribute re-attachment.** (a) x = statistic (four-cycle), y = z; planted four-cycle cluster at z = 239.43 against the degree-matched control at z = −0.84; unit of analysis = one synthetic network per bar. (b) x = per-instance attribute value in attribute units, y = frequency; full distributions of two size-3 classes with similar central summaries but different instance counts, 2,838 against 11,710, planted maximum 6.48 and comparison-class maximum 0.38 marked. Both panels are synthetic: the real data enter this article only as the published per-class aggregates of Table 5, so no real-data panel is shown and no per-instance distribution is displayed for either source. Error representation: (a) z against the null ensemble; (b) full empirical distributions rather than summary intervals. Takeaway: the pipeline detects planted structure when it exists, and the attribute step is demonstrated on the synthetic panel only.

Supports: the pipeline passes its positive and negative controls; re-attachment is reproducible and is demonstrated on the synthetic panel. Does not support: attribute re-attachment as a method with information gain, any attribute-adjusted inference, or any statement about the per-instance attribute distributions of either real data source, which were not accessed.

## Negative results and the pre-interpretation check

Four results above are negative or corrective and all four are in the main text: fan-class significance is undefined under the stated null (Fig. 1), the raw fan counts and the size-3 ratio in the cross-dataset contrast carry nothing beyond the degree sequences (Fig. 2), the point decomposition of the open-path deficit does not identify which component carries it and does not hold up under leave-one-tourist-out deletion or a change in the edge rule (Fig. 4, Fig. 5), and attribute re-attachment shows no gain that these data could carry (Fig. 7); one check changed the numbers, the sampler counting rule (Fig. 6). Three limits are visible in the numbers themselves: with 17 agents the tourism analysis is a worked example rather than a population estimate, with four classes the cross-dataset association cannot resolve below p = 0.0417, and removing two of the seventeen respondents moves the mixing share from 63.4 to 15.1 %. The steps that would have caught each of these before interpretation are collected in Fig. 8.

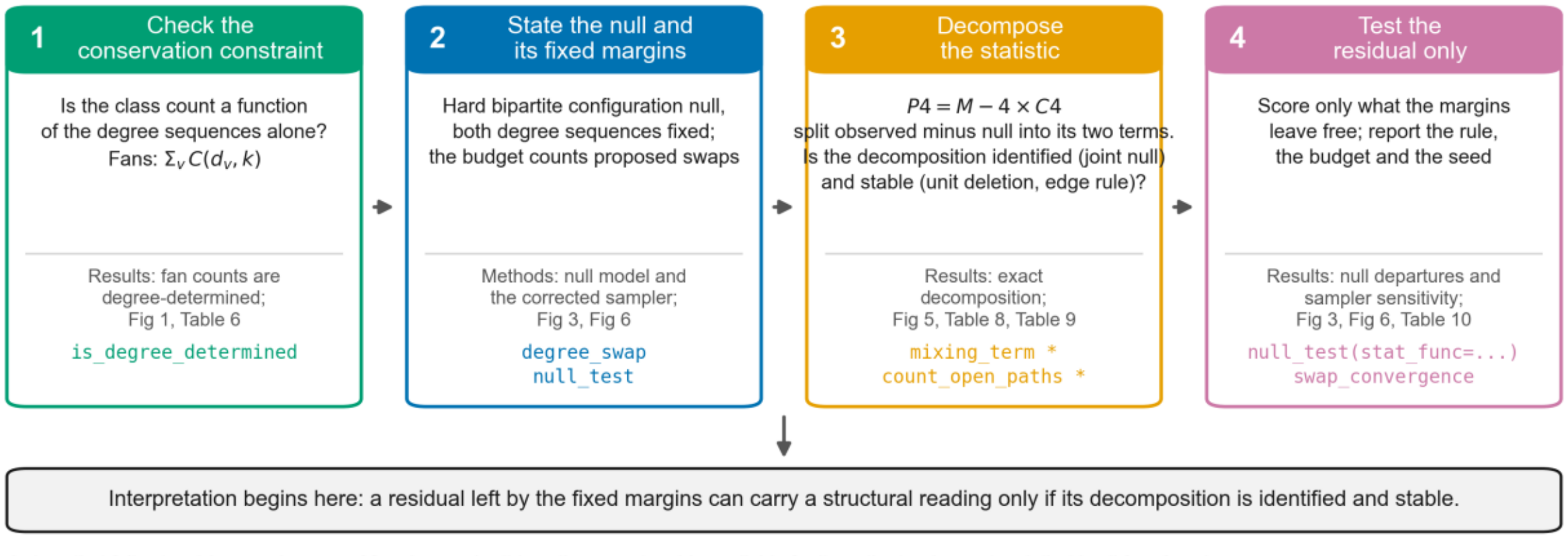

**Figure 8. Pre-interpretation check.** Four-step flow diagram, no axes: check the conservation constraint on the motif class, state the null and which margins it fixes, decompose the statistic into margin-determined and residual parts and ask whether that decomposition is identified under the joint null and stable under deletion and under the edge rule, test the residual only. Unit of analysis = one motif statistic passing through the audit. Each step links to the corresponding identity, the results subsection applying it, and the code entry point implementing it; no error representation, because the figure is a procedure rather than a measurement. Takeaway: interpretation begins after, not before, the deterministic component is removed.

Supports: an executable checklist grounded in the identities and results above. Does not support: any claim that the checklist has been validated prospectively.

## Discussion

### Answering the research question

The answer to the research question is asymmetric. Fan-motif counts contain no residual variation under a hard null that fixes both bipartite degree sequences, so their null-model significance cannot support a structural interpretation. The tourism open-path count is not similarly conserved and shows a deficit of −2,243 instances, 2.4 % of the null mean, under the corrected sampler, and that deficit is robust to which tourist is dropped (Table 8). Its exact decomposition splits the deficit into a mixing component and a four-cycle component, but that split does not assign the deficit to either. The two components are almost collinear under the null, the observed pair lies outside the region of the joint null the ensemble actually samples, and the split moves across the whole plausible range under leave-one-tourist-out deletion (Fig. 5, Table 8). The deficit is therefore extreme relative to the sampled hard-degree null for this observed graph, while its class-level interpretation is undetermined and unstable.

Taken sub-question by sub-question: fan significance is undefined by construction rather than merely weak (Fig. 1, Table 6); in the tourism network the raw fan counts, and the size-3 fan-out to fan-in ratio, add nothing the two degree sequences do not already state, while on the luxury side the same reading would be an inference from the counting identity rather than a measurement, because no edge list exists there and the published counts do not satisfy the bound that identity implies, so no verdict is drawn on that side; the size-4 within-size shares are not degree-determined and are reported descriptively (Fig. 2, Table 5); of the size-4 statistics that are free to move, the open path and the mixing term depart from the null while the four-cycle count does not, and the joint position of the pair lies outside the sampled null cloud without identifying which of the two carries the departure (Fig. 3, Fig. 5, Table 7); and attribute re-attachment reproduces its workflow on the synthetic panel while no gain beyond typed enumeration can be shown on the real side, where only published per-class aggregates are available (Fig. 7).

### What the residual on the tourism network is

What the surviving departure is cannot be settled from these data. The identity splits the open-path count into an edge-wise degree product and a four-cycle correction. Marginally the mixing term is low, below all 500 samples drawn, and the observed edge degree assortativity is more negative than the ensemble allows, while the four-cycle count sits inside its null range. The joint null does not resolve the split any further. The two terms are almost collinear under it ($r = 0.968$), and the observed mixing term lies outside the range the ensemble samples, so the null carries no conditional distribution at the observed value and a conditional expectation formed there would be an extrapolation rather than a null expectation; what the ensemble does show is that the observed pair sits far from the null cloud as a whole. A single point split cannot decide between the two components, and these data supply no third quantity that would. We therefore state what the analysis establishes, that the open-path count is deficient under this null on this graph, and we do not state which term carries it.

Two perturbations reinforce that restraint. Removing the two most active tourists leaves a graph on which the deficit sits with the four-cycle term instead (Table 8), and raising the rating threshold that defines an edge removes the deficit altogether (Table 9). The threshold is therefore not a nuisance choice that leaves the result standing; it is a definitional choice the result depends on, and the deficit is a property of the rating $> 0$ graph. Popularity-driven visiting and the sampling of the original 17 respondents remain compatible with the observation and cannot be separated from each other on one dataset. The rating threshold, by contrast, is separable, has been separated, and the deficit does not survive it.

### Relation to the nearest prior work

We claim no theorem. The conservation result for stars under the configuration model, including the k = 2 and k = 3 cases used here, is already in the literature (Wegner, 2014). What this paper adds is the demonstration, on an applied comparison of our own, that the raw fan counts and the size-3 fan-out to fan-in ratio are absorbed by that result, together with the exact decomposition of the one statistic that is not and the finding that the decomposition does not identify an attribution (Tables 1 to 3, Table 8, Table 9, Fig. 5). Item by item: to the tourism study (Shao, Ieiri and Hishiyama, 2021) we add the degree-preserving null and the determinacy verdict and change nothing about its published results; to the luxury studies (Shao, Ieiri and Takahashi, 2025a; Shao, Ieiri and Takahashi, 2025b) a comparability warning on published aggregates we could not go behind, together with the observation that their published item-side counts do not meet the minimum edge count the same identity implies for the transaction total reported alongside them, which is why that side is carried here as an illustration and not as a second worked dataset. Relative to the cross-domain comparison (Shao, Ieiri and Hishiyama, 2024) we withdraw the structural reading of the fan contrast that the withdrawn manuscript drew from it, and replace it with the degree-sequence account. Relative to our own software (Shao, 2026), whose *is_degree_determined* guard registers exactly the four fan classes, this paper is the worked example behind that guard and covers what it does not: the open path and the mixing term, scored through a custom statistic, and their decomposition. The statistical and software work reviewed in Tables 1 to 3 supplies the null families and the enumeration; none of it is a counter-example to the point made here.

### What judgement this changes

The practical change is one of ordering: check determinacy, then choose a null, then decompose, and only then test. A motif table that reports fan classes together with p-values from a hard degree-preserving null is not a weak result to be read with caution: those p-values carry no information about the network beyond its degree sequences, and the same holds for any comparison of such rows across networks or across time. A second failure mode sits beside that one, and it is the one our own luxury analysis committed: screening the same classes against an edge-count-preserving baseline that does not preserve degrees produces enrichment rather than zero variance, because the observed degree sequence is heterogeneous and the baseline's is not, so a fan class can pass a significance filter on the strength of exactly the degree information the baseline discarded. The two failures look opposite, a degenerate reference distribution against a manufactured one, and the first step of the checklist covers both, because it asks what the degree sequences already fix before any null is chosen. Conversely, a statistic that is not degree-determined should be decomposed before it is interpreted, and the decomposition should then be checked for identifiability and for stability before it is read as an attribution. Here the point split assigns most of the departure to the edge-wise degree product, and neither the joint null distribution of the two components nor the deletion and threshold re-runs support that assignment. This is a check we adopted for our own re-analysis, not a protocol that has been validated prospectively; its value is that each of its four steps is decidable from the data at hand (Fig. 8).

### Who is affected, and what they would do differently

Three groups of readers face a concrete decision. For an applied researcher, the check costs one function call, and its outcome determines whether a null-model column belongs in the table at all. For a reviewer or a reproducibility auditor, it is a single question to ask of any bipartite motif table: which of these classes can move once both degree sequences are fixed? Rows that cannot move should be read as degree summaries and compared as such. For a maintainer of motif software, the consequence is a guard in the counting layer plus an explicit statement of the swap-counting rule; our own package carries the first and, since version 1.1.1, the second. We do not claim that adopting this ordering changes published conclusions in any particular literature; that would require a survey of applied motif reporting, which we have not done.

### Boundaries, failure conditions, and alternative explanations

The determinacy result is specific to nulls that fix both degree sequences exactly; a null that fixes only expected degrees is a different estimand and is not examined here (Squartini and Garlaschelli, 2011; Saracco et al., 2015; Neal et al., 2024). Under such a family the fan counts do acquire a distribution, and nothing in our evidence says what that distribution looks like on these networks. The result is also relative to the typing of nodes and instances fixed in Methods. Three failure conditions bear on the residual reading rather than on the determinacy reading, and the first of them has been triggered. It was stated as follows: if the deficit were carried by the four-cycle term instead of the mixing term, the point split would not support an attribution to degree mixing. Removing the two most active tourists produces exactly that configuration, with a mixing z of −1.2 against a four-cycle z of +3.8 and a mixing share of 15.1 %, on a graph of 485 edges. The condition is met, and the attribution is withdrawn accordingly; what remains is the deficit itself. Second, if the null ensemble were mis-sampled, the reference distribution would be wrong; we therefore report z-scores and ensemble quantiles rather than an empirical p-value at the resolution floor of the sample, and we repeated the analysis across four chain lengths, two seeds and two implementations by the author, a local script and the attrimotif package, which agree on the sign, the magnitude, and the share (Fig. 6, Table 10). Third, the deficit is a property of one observed graph under one edge rule; a second tourism network could behave differently, and the same network at higher rating thresholds does behave differently (Table 9); nothing here is a population statement.

### Limitations

1. No attribute-preserving null exists in our implementation, so the per-class attribute distribution D(c) remains descriptive and no instance-level significance is reported; the re-attachment step is demonstrated on the synthetic panel only, since no per-instance attribute data were accessed for either real source (N1).

2. The luxury aggregates provide no edge list, so no edge-level null can be run on them. They are also not reconcilable with the transaction total published beside them: the item-side counts $2.23\times10^6$ and $3.63\times10^7$ require at least 89,502 customer-item edges, against 26,451 reported transactions for the same 36 networks, and this note cannot decide from published figures whether multi-edges, pooling across months or a different edge definition in the source accounts for the gap. The luxury side is therefore a provenance-limited illustration of the counting identity and carries no determinacy verdict of its own (N2).

3. The luxury profile sums 36 independently degree-sequenced monthly networks while the tourism network is a single static snapshot; motif counts do not aggregate linearly, and the per-month node and edge cardinalities are not reported in the source, so the two sides are not units of the same kind (N3).

4. With two datasets, domain effects and dataset effects cannot be separated, and no cross-domain statement is made (N4).

5. The tourism network has 17 agents; the counts are exact for that graph, but the results are a worked example and not a population estimate, and discreteness at this size is a reason to read the ensemble quantiles alongside the z-scores (N5).

6. The numbers reported here come from a proposal-counted sampler: the version of the public code deposit published before this study (v4) shipped the accepted-count rule, and version 5, which carries the proposal-counted sampler, is published under that same concept DOI, with version DOI 10.5281/zenodo.22287908; measured against the four proposal-counted chain-length ensembles the difference on this network is at most 0.27 in z, whether taken against the record run or across all four chain lengths, and it reaches 0.31 against the older local re-runs and 0.35 against the corrected deposit script (Results); it moves the mixing share from 66.2 % to 64.4 %. All of these are smaller than the 0.66 seed-to-seed difference at one chain length, computed from the unrounded z values (rounded values give 0.67), so the sampler correction is a reporting correction and not the source of the instability described in Limitations 8 and 9 (N6). The chain length of record was chosen after all four ensembles had been computed, as a reporting convention; a principled stopping rule for randomly sampling bipartite networks with fixed degree sequences was available (Neal, 2025) and was not applied here, so the ensemble length reported above is not certified as sufficient by any pre-specified criterion.

7. Only one null family, the hard configuration model, is used; soft-margin families are left to future work.

8. The point decomposition of the open-path deficit does not identify which component carries it. Under the null the mixing term and the four-cycle count correlate at 0.968, so the two coordinates cannot be moved independently within the null cloud; and the observed mixing term lies below all 500 samples drawn, so the ensemble supplies no conditional distribution at the observed value and no conditional null expectation is available there. Reading the joint null further than a distance would therefore require a conditional sampler that these ensembles do not provide, and no quantity in these data separates the two components. The 64.4 % mixing share is an accounting figure and is not reported as an attribution.

9. The open-path deficit is a property of the graph defined by an edge at rating > 0. It survives every leave-one-tourist-out re-run and the removal of the two most active tourists together, but the decomposition does not (Table 8), and the deficit itself does not survive a higher rating threshold (Table 9), where power and the identity of the analysed graph change together and cannot be separated on one dataset. With 17 respondents, two of whom rated 76 of the 80 sites each, no result here should be read as a property of the destination rather than of this sample under this edge rule.

10. The joint-null, deletion and threshold analyses were not part of the analysis plan; they were added once the point split was in hand, and are reported as such.

## Conclusions

Bipartite fan-motif counts should not be interpreted as residual local structure under a null that already fixes the degree sequences determining those counts. In the tourism network analysed here, the raw fan counts and the size-3 fan-out to fan-in ratio are a degree-sequence signature, while the within-size shares at size 4 are not degree-determined and are reported descriptively. The published luxury aggregates are carried alongside as an illustration only: their counts do not satisfy the edge bound the same identity implies, so no determinacy verdict is drawn on that side and their 99.8 % fan-in stands as a published descriptive value. In the reconstructable tourism network, the statistic that remains, the open-path deficit, is extreme relative to the sampled hard-degree null for this observed graph and survives every leave-one-tourist-out re-run; it can be decomposed exactly, but that decomposition is not an attribution. Its two components are almost collinear under the null, the observed pair lies outside the region of the joint null the ensemble samples, so no conditional null expectation is available at the observed value, the split is unstable under leave-one-tourist-out deletion, and the deficit itself does not hold under a higher rating threshold. The deficit is therefore extreme relative to that sampled null, while its class-level interpretation is undetermined and unstable; the value of the audit is that it shows this before an interpretation is published. The practical contribution is a pre-interpretation check: identify what is conserved, state the constrained margins, decompose the statistic and ask whether that decomposition is identified and stable, and test only what remains. The code, the anonymised tourism matrix, and the swap-counting rule behind every number above are stated so that each step can be recomputed. We claim no theorem, no generalisation beyond this one network and the published aggregates illustrated alongside it, and no domain-level regularity.

## Acknowledgements

The author thanks the authors of the data-collection study, Yuya Ieiri, Yuki Nakajima and Reiko Hishiyama, for agreeing to the public deposit of the de-identified rating matrix.

## Additional Information and Declarations

### Data availability

The anonymised tourism rating matrix (17 tourists x 80 sites, edge list T1-T17, no personally identifying information) is archived on Zenodo under CC BY 4.0 at 10.5281/zenodo.22299150 (README, matrix and licence). The matrix was collected in a Kyoto experiment with 17 tourists using a walk-rally application and reported in Ieiri, Nakajima and Hishiyama (2018), IEICE Transactions on Information and Systems (Japanese edition) J101-D(9):1325-1333 (in Japanese) (Ieiri, Nakajima and Hishiyama, 2018), which the 2021 study cites as its reference [23] for the origin of the data; the present note is, like that study, a secondary user of it. A rating of 0 in the matrix means the spot was not visited and no evaluation was recorded. The consent obtained in the data-collection study (Ieiri, Nakajima and Hishiyama 2018) covers public re-use of the de-identified matrix, as confirmed by the author with that study's authors; access does not depend on contacting the author. The luxury motif-instance aggregates reported here are taken from the published sources (JIP 2025); the underlying customer-item transaction records are proprietary to a commercial partner and were not accessed for this study. No new luxury data are released.

### Code availability

The predecessor framework package, which carries the corrected proposal-counted sampler and the synthetic validation script, is deposited on Zenodo under the concept DOI 10.5281/zenodo.21122972. The version published before this study is v4 (DOI 10.5281/zenodo.21441291; earlier versions were withdrawn by the author) and it shipped the accepted-count rule; version 5, which carries the proposal-counted sampler behind the sampler numbers reported here, is published under that same concept DOI, with version DOI 10.5281/zenodo.22287908. That package predates this article and does not contain its analysis. The scripts, cached null ensembles and figure code of this article, which reproduce every number reported above from the deposited matrix, are archived separately at 10.5281/zenodo.22308052. Degree-preserving nulls were computed with attrimotif 1.1.1, Zenodo DOI 10.5281/zenodo.22259239 (software described in a companion SoftwareX manuscript, currently under review).

### Ethics

The data-collection study (Ieiri, Nakajima and Hishiyama 2018) did not require ethics review, as confirmed by Waseda University at the author's enquiry, and no new human-subject data were collected for this note; it reanalyses an anonymised rating matrix from a prior study and published aggregate statistics. The de-identified matrix is released under the consent obtained in the data-collection study (Ieiri, Nakajima and Hishiyama 2018), which covers its public re-use; no new consent procedure applied to this note.

## Funding

This work was supported by the Waseda University Grant for Special Research Projects. This work was supported by MEXT Supporting Pioneering Research through AI for 1,000 Discovery challenges Program (SPReAD) Japan Grant Number JPMXP1726306059.

## Generative AI use

The full statement of generative AI use, naming the tool and its version, how it was used and why, is given in Materials and Methods under "Use of generative AI".

## Competing interests

The author develops and maintains the attrimotif package used for the null models; a description of the package is under review at SoftwareX (unpublished). The author declares no other competing interests.

## Author contributions

Tengfei Shao is the sole author and is responsible for conceptualisation, methodology, software, formal analysis, data curation, writing, and visualisation.

## Prior related work disclosure

This submission reanalyses the tourism rating matrix collected in the Kyoto walk-rally experiment reported by Ieiri, Nakajima and Hishiyama (IEICE Transactions D, 2018) and reused in the author's co-authored tourism study (IEICE 2021), together with the per-class motif aggregates published in the luxury studies (JIP 2025, x2), and it re-reads the cross-dataset demonstration of JIP 2024. The size-3 fan-out to fan-in profile of the tourism network and the normalised-entropy comparison between the two profiles were not reported in IEICE 2021 or JIP 2024: they were computed in an earlier manuscript submitted to Scientific Reports, which used overlapping data and code and advanced the structural and cross-domain interpretations that are withdrawn here. That manuscript is therefore the locus of the withdrawn interpretation; it was rejected after review and has never been posted as a preprint. The attrimotif package, version 1.1.1, is cited here as software for the sampler and the degree-determinacy guard, and a description of the package is under review at SoftwareX (unpublished); the present contribution is the worked example and the open-path decomposition, which that software does not provide.